\documentclass[a4paper,fleqn,usenatbib]{mnras}

\usepackage[T1]{fontenc}
\usepackage{ae,aecompl}
\usepackage{hyperref}
\usepackage{breakurl}

\newcommand{\jb}{\textit{J\/}}
\newcommand{\h}{\textit{H\/}}
\newcommand{\ks}{\textit{K\/}$_\mathrm{s}$}

\newcommand{\gaia}{\textit{Gaia\/}}

\newcommand{\spitzer}{\textit{Spitzer\/}}

\newcommand{\wise}{\textit{WISE\/}}

\newcommand{\planck}{\textit{Planck\/}}
\newcommand{\msun}{\,M$_{\sun}$}

\newcommand{\av}{\textit{A\/}$_\mathrm{V}$}

\newcommand{\kms}{km\,s$^{-1}$}

\usepackage{url}

\usepackage{graphicx}	% Including figure files
\usepackage{amsmath}	% Advanced maths commands
\usepackage{amssymb}	% Extra maths symbols
\usepackage{enumitem}
\usepackage{pdflscape}

\title[The \textit{Gaia} view of the Cepheus flare]{The \textit{Gaia} view of the Cepheus flare}

\author[Szil\'agyi  et al.]{M\'at\'e Szil\'agyi$^{1,2}$\thanks{E-mail: szilagyi.mate@csfk.org}, M\'aria Kun$^{1}$,
P\'eter \'Abrah\'am$^{1,3}$ \\
$^{1}$Konkoly Observatory, Research Centre for Astronomy and Earth Sciences, E\"otv\"os Lor\'and Research Network (ELKH),\\ 
H-1121 Budapest, Konkoly Thege Mikl\'os \'ut 15--17, Hungary\\
$^{2}$Department of Astronomy, E\"otv\"os Lor\'and University, H-1117 Budapest, P\'azm\'any P\'eter s\'et\'any 1/A, Hungary\\
$^{3}$ELTE E\"otv\"os Lor\'and University, Institute of Physics, P\'azm\'any P\'eter s\'et\'any 1/A, 1117 Budapest, Hungary}

\date{Accepted XXX. Received YYY; in original form ZZZ}

\pubyear{2021}

\begin{document}
\label{firstpage}
\maketitle

\begin{abstract}
We present a new census of candidate pre-main-sequence stars in the Cepheus flare star-forming region, based on \gaia~EDR3 parallaxes, proper motions, and colour--magnitude diagrams. We identified new candidate members of the previously known young stellar groups associated with NGC\,7023, L1177, L1217/L1219, L1228, L1235, and L1251. We studied the 3D structure of the star-forming complex and the distribution of tangential velocities of the young stars. The young stellar groups are located between 330 and 368\,pc from the Sun, divide into three kinematic subgroups, and have ages between 1--5 million years. The results confirm the scenario of propagating star formation, suggested by previous studies. In addition to the bulk pre-main-sequence star population between 330 and 370~pc, there is a scattered and more evolved pre-main-sequence population around 150--180~pc. We found new candidate members of the nearby Cepheus Association, and identified a new moving group of 46, 15--20 million years old pre-main-sequence stars located at a distance of 178~pc, around the A0-type star HD\,190833. A few pre-main-sequence stars are located at 800--900~pc, indicative of star-forming regions associated with the Galactic local arm above the Galactic latitude of +10\degr.
\end{abstract}


%%%%%%%%%%%%%%%%%%%%%%%%%%%%%%%%%%%%%%%%%%%%%%%%%%

%%%%%%%%%%%%%%%%% BODY OF PAPER %%%%%%%%%%%%%%%%%%

\section{Introduction}
It was noticed by \citeauthor{Hubble1934} in \citeyear{Hubble1934} that the opaque belt of the Galactic plane extends to latitudes as high as +20\degr\ in the Cepheus, between the Galactic longitudes 93\degr\ and 133\degr. Observations of the atomic and molecular interstellar matter  \citep[][respectively]{Heiles1967,Grenier1989} have shown that the obscuring dust belong to two giant interstellar clouds, containing a total mass of some $1.3\times10^{5}$\,\msun, and separated from each other in radial velocity. The high-velocity cloud in the 100\degr < \textit{l} < 140\degr, +10\degr < \textit{b} < +17\degr\ region and in the $-$20\,\kms < \textit{v}$_\mathrm{LSR}$ < $-$8\,\kms\ interval is a high-latitude extension of the Local arm of the Galaxy around 800--900\,pc. The low-velocity cloud at 100\degr < \textit{l} < 115\degr, +10\degr < \textit{b} < +20\degr, and in the radial velocity interval $-$8\,\kms < \textit{v}$_\mathrm{LSR}$ < +8\,\kms\ is a nearby giant molecular complex around a distance of 300\,pc, one of the nearest star-forming regions of the northern sky. On the high-longitude side it is bordered by a region free of interstellar matter and radiating in soft X-rays, suggesting a hot bubble, created by a supernova \citep{Grenier1989}. The presence of an expanding shell, named \textsl{Cepheus Flare Shell, CFS\/}, centred on the hot bubble, was identified in the distribution of the neutral hydrogen by \citet*{Olano2006}. The CFS might have been created by the winds and subsequent explosion of a high-mass star some 5~million years ago. The giant radio continuum structure \textsl{Loop~III\/}, identified by \citet{Berk1973} and discernible also in the \textit{WMAP\/} K-band polarization map \citep{Page2007} is nearly concentric with the CFS.

The molecular gas of the region was mapped in the $^{13}$CO line by \citet{YDM1997}. The clouds are distributed over a wide range of radial velocities. Several authors argue that this may reflect multiple supernova events in the past  \citep{Grenier1989,Bally2001,Olano2006}, which might have affected the star formation in this region.

Low-mass young stellar objects (YSOs) of the Cepheus flare are clustered on a few molecular clouds, scattered over the whole surface of the complex \citep*{kun1998,kun2008}. Ongoing star formation can be observed in the L1148/L1158 complex, L1172/L1174 (associated with NGC\,7023), L1177 (CB~230), L1219, L1228, L1251, L1262 (CB~244) \citep[see][and references therein]{Froebrich2005,Connelley2008,Kirk2009}.
A less studied star-forming cloud is L1235, associated with a few optically visible low-mass pre-main-sequence (PMS) stars \citep{kun2009} and submillimeter sources \citep{scuba}. The small dark cloud Lynds~1221, located near the low-latitude boundary of the region is associated with only deeply embedded young stellar objects (YSOs) \citep{Young2009}, therefore its distance is uncertain. The nearby \textsl{Cepheus association} of 20--30~Myr old PMS stars, located at $157\pm10$\,pc  \citep{Klutsch2020} is projected at the high Galactic longitude side the Cepheus flare. The very active low- and intermediate-mass star-forming region NGC\,7129, located at the low-latitude boundary of the region at a distance of 880~pc \citep{Reid2014}, is probably associated with the \textsl{Cepheus Bubble\/}, blown by the hot, high-mass stars of the Cepheus\,OB2 association  \citep{Kun1987,Abraham2000}.

Lists of spectroscopically confirmed PMS stars in the Cepheus flare region were presented by \citet{kun2009}.  \citet{Kirk2009} published results of the \spitzer\ survey of the most prominent star forming clouds of the Cepheus flare. Most of the stars in both lists are disc-bearing young stars \citep[Class\,II YSOs, or classical T~Tauri stars, CTTS,][]{greene1994}. Discless PMS stars  (Class\,III YSOs or weak-line T~Tauri stars, WTTS), distributed over the surface of the Cepheus flare, were identified by \citet{TNK2005}.  

The cloud complex extends over some 15~degrees in Galactic longitude, corresponding to some 90\,pc at the mean distance of 350\,pc \citep{Dzib2018}. This size suggests a significant depth of the cloud complex along the line of sight. 
The precise positions, parallaxes, proper motions, and photometry of the optically visible young stars, measured by the \gaia\ mission \citep{gaia2016b}, opened the door to studying of the spatial structure and internal motions of star-forming regions.

The overall properties of the Cepheus flare, based on \gaia~DR2 \citep{gaiadr2} data of 47 pre-main-sequence (PMS) stars were investigated by \citet{Dzib2018} during the study of the \gaia\ view of the Gould Belt system of the nearest star-forming regions. They found that the average distance of the known young stars of the Cepheus flare, included in the \gaia~DR2, was $360\pm32$~pc. They also found that the proper motion dispersions of the studied stars were large compared to those of other regions.

\citet{green2019} presented three-dimensional maps of Galactic dust reddening, based on \gaia\ parallaxes and stellar photometry from Pan-STARRS~1 and 2MASS. The maps suggest that the clouds of the Cepheus flare are located between 320 and 370~pc, and toward a few lines of sight there is a more distant layer of dark clouds between 850--1000~pc.

\citet{Zari2018} selected pre-main-sequence stars and young upper main sequence stars in the solar neighbourhood from the \gaia\ DR2 based on a combination of astrometric and photometric criteria. Their catalogue of pre-main-sequence stars contains 826 stars in the $100\degr \leq l \leq 124\degr$, $8\degr \leq b \leq 24\degr$ area and closer than 500~pc to the Sun. This sample is biassed against the youngest members of the PMS population, whose photometric properties are affected by the circumstellar disc and envelope. 

The recently released \gaia~EDR3 \citep{Gaia2020} represents a significant improvement with respect to \gaia~DR2 in terms of astrometric and photometric precision, accuracy, and homogeneity. In this paper we present a view of the Cepheus flare based on \gaia~EDR3 data.  We examine the star-forming molecular clouds of the Cepheus flare one by one. We derive mean distances and tangential velocities of the young stellar groups associated with individual clouds, and identify new candidate members.  To characterize the new candidate young stars we examine their \textit{G} vs. (\textit{G}$-$\textit{G}$_{\mathrm{RP}}$) colour--magnitude diagram, \textit{2MASS\/} and \wise\ colour--colour diagrams, and estimate mean ages using the PARSEC \citep{Bressan2012} isochrones for stars more massive than 1.4\,\msun\ and CIFIST \citep{BHAC2015} for the stars below this mass. We present our initial sample of young stars in Sect.~\ref{sect2}. Our search for new members is described in  Sect.~\ref{sect3}, and the results are  presented and discussed in Sects.~\ref{sect4} and \ref{sect5}. We summarize the main results in Sect.~\ref{sect6}.

\section{Distances and tangential velocities of individual young stellar groups}
\label{sect2}

In order to determine the mean distances and the range of proper motions of young stellar objects associated with individual molecular clouds we compiled an initial list of \gaia~EDR3 counterparts of known, optically visible YSOs in the $100\degr < l < 125\degr$, $+8\degr< b < +22\degr$\ region, consisting of  spectroscopically identified YSOs published in \citet{kun2009}, 
\spitzer\ sources, identified as flat-SED and Class\,II YSOs by \citet{Kirk2009},
spectroscopically identified weak-line T~Tauri stars from \citet{TNK2005}, and spectroscopically identified WTTS members of the \textit{Cepheus Association\/} from \citet{Klutsch2020}.

\begin{figure*}
\centering \includegraphics[width=16cm]{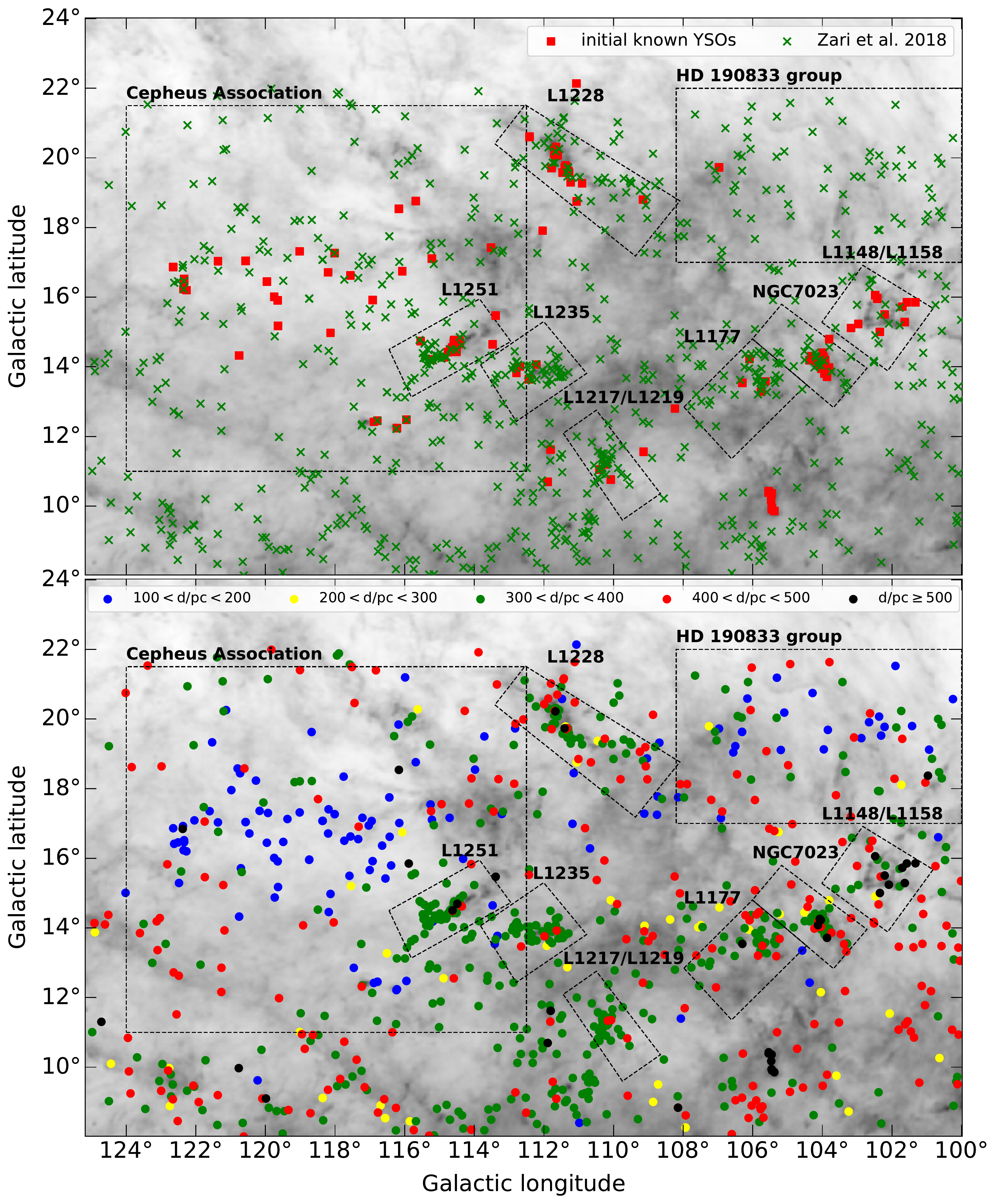}
\caption{\textit{Top\/}: Distribution of spectroscopically confirmed young stars and \spitzer\ sources (red squares) from our initial list and candidate pre-main-sequence stars of \citet{Zari2018} (green crosses) in Galactic coordinates, overplotted on the \planck~857~GHz map of the region.
\textit{Bottom\/}: Same as the upper panel, but the symbol colours indicate distances.}
\label{fig1}
\end{figure*}

We searched \gaia~EDR3 counterparts of these stars within 1\,arcsec. Nine of the 77 pre-main-sequence stars listed in \citet{kun2009} are  associated with NGC\,7129 at a distance of 880\,pc, and three further ones (2MASS J21225426+6921344, [KBK2009b]\,Em*\,119\,S, and TYC 4608-2063-1) are located far beyond the molecular complex. Based on the angular separations and position angles listed in \citet{kun2009} we could identify both components of the visual binary T~Tauri stars Cl*\,NGC 7023\,RS\,5, LkH$\alpha$\,428, OSHA\,48, OSHA\,50, and [K98c]\,EM*119. Sixty-six infrared sources, classified as YSOs in \citet{Kirk2009}, and missing from the above list, coincide with \gaia~EDR3 sources within 1\,arcsec. Sixteen  weak-line T~Tauri stars, resulted from spectroscopic observations of \textit{ROSAT\/} X-ray sources in \citet{TNK2005}, have \gaia~EDR3 counterparts within 1\,arcsec. These stars are projected outside of the molecular clouds and are located at various distances. Twenty-nine pre-main-sequence stars were identified by \citet{Klutsch2020} as members of the nearby \textit{Cepheus Association\/}. Each of them have \gaia~EDR3 counterpart, and seven of them coincide with weak-line T~Tauri stars identified by \citet{TNK2005}. We used four additional stars from \citet{Faherty2018} which appear in \citet{Klutsch2020} as a \textit{Cepheus Association\/} bona-field members. After removing the duplicates, our initial list contains 176 known young stars. 
We supplemented the data in our initial list with distances, derived by \citet{BJ2021} from \gaia~EDR3 parallaxes, using a prior constructed from a three-dimensional model of our Galaxy, which takes into account the interstellar extinction and \gaia's variable magnitude limit. Using {\fontfamily{pcr}\selectfont
Astropy} \citep{astropyI} we transformed the proper motion components into Galactic proper motions, and using the distances the Galactic tangential velocity components were calculated. 
The initial list of young stars with \gaia~EDR3 counterparts is shown in Table~\ref{tab:yso_initial}. Their distribution in Galactic coordinates is plotted with red squares in the upper panel of Fig.~\ref{fig1}. Green crosses in the same panel show candidate pre-main-sequence stars from \citet{Zari2018}. Symbol colours in the lower panel indicate distance intervals.

Then we  defined areas of the sky containing the clouds and their associated YSOs. To include possible scattered young stellar populations outside of the visible boundaries of the clouds we defined wide tetragons encompassing the clouds. The tetragons are shown in Fig~\ref{fig1}. In addition to the known star-forming molecular clouds we defined two large regions containing numerous pre-main-sequence stars and candidates closer than 200\,pc (\textit{Cepheus Association\/} and \textit{HD\,190833 Group\/}, see Section \ref{sect:comoving}). We suppose that the candidate pre-main-sequence stars from \citet{Zari2018},  projected within our tetragons, are members of the YSO clusters and use them to derive the mean distance and tangential velocity components of the clusters. 

\begin{figure*}
\centering \includegraphics[width=16cm]{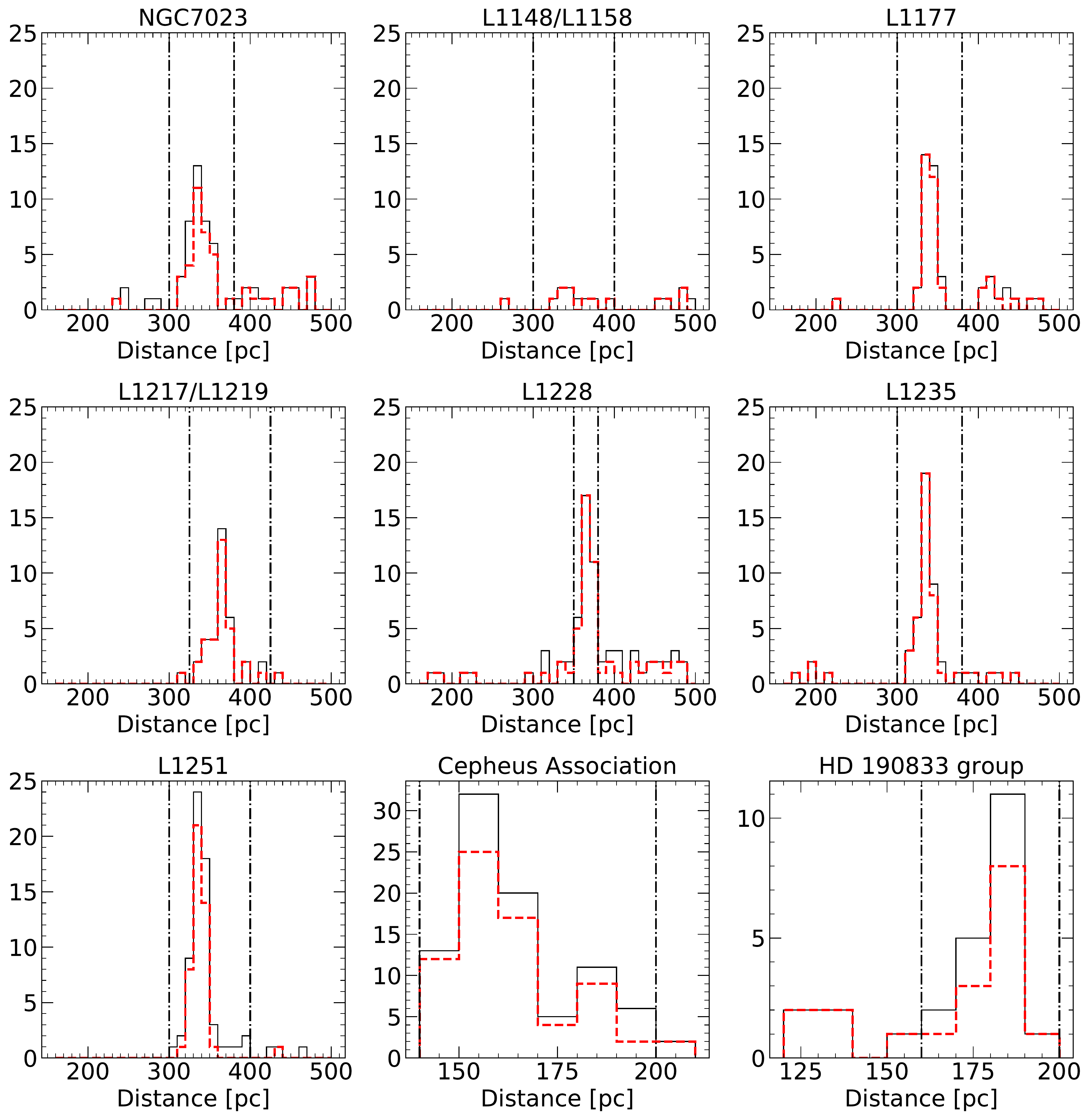}
\caption{Distance histograms of the YSOs of our initial list supplemented with candidates from \citet{Zari2018}. The thin black lines indicate all stars, and the thick red dashed lines show stars with $\varpi/\sigma_\varpi > 10$, $|\mu_{\alpha}^\star/\sigma_{\mu_{\alpha}^\star}|$ > 2, $|\mu_{\delta}/\sigma_{\mu_{\delta}}|$ > 2 and $\mathrm{RUWE}<1.6$. The sources outside the region bounded by vertical dashed-dotted lines are considered fore- or background stars and were disregarded. For plotting purposes we did not plotted stars with distances larger than 500\,pc.}
\label{fig:disthist}
\end{figure*}

We plotted in Fig.\,\ref{fig:disthist} the distance histogram of the YSOs listed in Table~\ref{tab:yso_initial} and candidates from \citet{Zari2018} for each region labeled in Fig.~\ref{fig1}, and identified the obviously foreground and background stars of the area. After removing these stars from the lists we calculated the mean parallax and proper motion components of the stars for each region.
During this process we used only stars which fulfilled the astrometric criteria as follows: $\varpi/\sigma_\varpi > 10$, $|\mu_{\alpha}^\star/\sigma_{\mu_{\alpha}^\star}|$ > 2, $|\mu_{\delta}/\sigma_{\mu_{\delta}}|$ > 2 and $\mathrm{RUWE}<1.6$, where $\varpi$ and $\sigma_\varpi$ are the parallax and parallax error, $\mu_{\alpha}^\star$, $\mu_{\delta}$ and $\sigma_{\mu_{\alpha}^\star}$, $\sigma_{\mu_{\delta}}$ are the proper motion components in ICRS ($\mu_{\alpha}^\star=\mu_{\alpha}cos\delta$), and their uncertainities, respectively, and RUWE is the re-normalized unit weight error, described in \citet{lindegren-ruwe}. Table~\ref{tab1} shows the central positions in Galactic coordinates, average parallaxes and proper motions of the YSOs of each cloud. The host molecular clouds, their distances, obtained from the 3D extinction maps of \citet{green2019}, and radial velocities, found in \citet{YDM1997} and \citet{CB1988} are also listed.

Based on the stars that fulfil our astrometric criteria we defined the distance and tangential velocity ranges of the confirmed and candidate YSOs for each region listed in Table~\ref{tab1}. First we calculated the median-absolute-deviation (MAD) of the distances for each region:
\begin{equation} \label{eq:mad}
    \mathrm{MAD(d) = median(|d - d_{med}|)},
\end{equation}
and adopted the distance range $\mathrm{d_{med}\pm5\,MAD(d)}$ as the line-of-sight extent of the cluster.  We examined the velocity ranges of the stars in these distance intervals and found that they are within $\mathrm{v_{l,med}\pm5\,MAD(v_{l}})$, $\mathrm{v_{b,med}\pm5\,MAD(v_{b}})$ for most of the clouds. An exception is L1228, where we used 3\,MAD for the distance and 4\,MAD for the velocities. The candidate PMS stars in the catalogue of \citet{Zari2018}, projected into the area of L1148/L1158 complex have a variety of motion directions. Apparently they do not constitute a comoving system. Therefore the tangential velocity ranges for this area were defined by the three stars listed in \citet{kun2009} and \citet{Kirk2009}. %(i) and (ii).
Table\,\ref{tab:filter} lists the boundaries of the tetragons and minimum and maximum values of the distances and velocities. 

We find that the YSOs, clustered on the dark clouds, are located between 300 and 400\,pc. A few T~Tauri stars %of samples (i)--(iii) 
of the initial list are situated beyond 500\,pc, indicative of distant star-forming regions around 800--1000\,pc. Stars with $100\,\mathrm{pc} < d < 200\,\mathrm{pc}$ are found in two large areas in the high-latitude part of the studied region.

\begin{table*}
\begin{center}	    	    
\caption{Central positions, mean parallaxes and proper motions of young stellar groups, and distances and radial velocities of their associated molecular clouds}	    	  
\label{tab1}	    	    
\begin{tabular}{lcccccccc} 
\hline	    	    
Name &	$l$ &  $b$ &  $\varpi$ & $\mu^{*}_{\alpha}$  & $\mu_{\delta}$  & \multicolumn{3}{c}{Molecular Cloud} \\
\cline{7-9}
&&&&&& Name & Distance$^*$ &  $v_\mathrm{LSR}$  \\
      & \multicolumn{2}{c}{(deg)} & (mas) & \multicolumn{2}{c}{(mas\,yr$^{-1}$)} & & (pc) & (km\,s$^{-1}$)  \\
\hline
 L1155 & 102.60 & 15.25 & 2.978 &   7.971 &  $-$1.699~ & [YDM97]\,CO\,8  & 341  & ~~2.9 \\ 
 L1174 & 104.06 & 14.19 & 2.943 &   7.480 &  $-$1.504~ & [YDM97]\,CO\,14 & 341  &  ~~2.7 \\ 
 L1177 & 105.17 & 13.16 & 2.903 &   8.041 &  $-$1.016~ &  CB 230  & 300  & ~~2.9 \\  
 L1219 & 110.59 & 12.07 & 2.605 &   5.055 &   1.439 & [YDM97]\,CO\,57 & 370  & $-$4.6 \\ 
 L1228 & 111.67 & 20.22 & 2.683 &   5.148 &   3.835 & [YDM97]\,CO\,66 & 381  & $-$7.6  \\ 
 L1235 & 112.24 & 13.89 & 2.908 &   5.400 &   1.649 & [YDM97]\,CO\,69 & 330  & $-$4.0  \\
 L1251 & 114.51 & 14.65 & 2.919 &   6.828 &   0.931 & [YDM97]\,CO\,79 & 341  & $-$3.8  \\
 L1262 & 117.20 & 12.36 & 6.346 &  21.647 &   1.409 & [YDM97]\,CO\,101& 210  & ~~3.9 \\
\hline
\end{tabular}
\end{center}
\smallskip
\flushleft{\small
$^*$Estimated on the 3D extinction map of \citet{green2019}.}
\end{table*}

\section{New members defined by distances and tangential velocities}
\label{sect3}  

\subsection{Selection criteria}
\label{sel_crit}

We search for further members of the star-forming clouds in \gaia~EDR3 based on the distances and tangential velocities, defined by the stars of our initial sample. First we selected all sources from \gaia~EDR3 with:
\begin{itemize}[align=left]
    \item[--] $100\degr < l < 125\degr$,
    \item[--] $8\degr < b < 22\degr$,
    \item[--] $d < 500\,\mathrm{pc}$,
    \item[--] $\varpi/\sigma_\varpi > 10$,
    \item[--] $|\mu_{\alpha}^\star/\sigma_{\mu_{\alpha}^\star}|$ > 2,
    \item[--] $|\mu_{\delta}/\sigma_{\mu_{\delta}}|$ > 2,
    \item[--] $\mathrm{RUWE} < 1.6$,
\end{itemize}
where $l$ and $b$ are the galactic longitudes and latitudes, $d$ is the distance (column r\_med\_geo) from \citet{BJ2021}. Using {\fontfamily{pcr}\selectfont
Astropy} we transformed the proper motion components into Galactic proper motions, and using the distances the Galactic tangential velocity components were calculated. 

We selected new candidate cluster members from the filtered \gaia~EDR3 list using the distance and tangential velocity ranges defined in Sect.~\ref{sect2}. 
No photometric constraint was applied during this process. The selection resulted in 266 new candidate members of the targeted clouds.  Figures~\ref{fig:ngc7023}a--\ref{fig:l1251}a show the surface distributions and tangential velocities of the cluster stars. Red symbols indicate stars of the initial list, green crosses mark the comoving members from \cite{Zari2018}, and blue symbols show the new candidate cluster members. Parallax histograms are plotted in Figs.~\ref{fig:ngc7023}c--\ref{fig:l1251}c. The colour--magnitude diagrams in Figs.~\ref{fig:ngc7023}e--\ref{fig:l1251}e support the  pre-main-sequence nature of the newly identified kinematic cluster members (see Sect.~\ref{sect3.3}).

\gaia~EDR3 identifiers, coordinates, distances, tangential velocities along Galactic longitude and latitude, and other names of these stars are listed in Table\,\ref{tab:yso-cand}. The number of confirmed YSOs with reliable \gaia~EDR3 data (N), the number of comoving candidate members (N$_\mathrm{cand}$), mean parallaxes  and tangential velocities of the clusters supplemented with the new candidate members are shown in Table~\ref{tab:result}.

\begin{table*}
\begin{center}	    	    
\caption{Boundaries defined for the clouds, minimum and maximum distances and tangential velocities of their known members.}	    	    
\label{tab:filter}	    	    
\begin{tabular}{lcccccccccc} 
\hline	    	    
Cloud &	$\mathrm{RA_{min}}$ & $\mathrm{RA_{max}}$ & $\mathrm{Dec_{min}}$ & $\mathrm{Dec_{max}}$ & $\mathrm{d_{min}}$ & $\mathrm{d_{max}}$ & $\mathrm{v_{l,min}}$ & $\mathrm{v_{l,max}}$ & $\mathrm{v_{b,min}}$ & $\mathrm{v_{b,max}}$   \\
      & \multicolumn{4}{c}{(deg)} & \multicolumn{2}{c}{(pc)} & \multicolumn{4}{c}{(km\,s$^{-1}$)}\\
\hline
NGC7023             &  313.75 &  317.50 &    67.00 &    70.00 &  301.506 &  372.512 &   3.246 &   8.206 & $-$12.759 &  $-$8.401 \\
L1148/L1158         &  307.50 &  313.00 &    66.50 &    68.80 &  327.065 &  332.835 &   4.423 &   5.622 & $-$12.286 & $-$11.713 \\
L1177               &  317.50 &  325.50 &    68.00 &    70.00 &  319.607 &  358.173 &   5.989 &   9.559 & $-$12.159 &  $-$9.272 \\
L1217/L1219         &  331.40 &  335.00 &    68.50 &    71.50 &  338.535 &  395.330 &   6.763 &  10.036 &  $-$5.532 &  $-$2.292 \\
L1228               &  311.25 &  318.50 &    74.00 &    79.00 &  355.051 &  380.205 &   8.580 &  13.183 &  $-$7.716 &  $-$0.578 \\
L1235               &  330.00 &  338.00 &    72.50 &    74.41 &  317.331 &  357.335 &   6.685 &  10.370 &  $-$4.139 &  $-$1.653 \\
L1251               &  334.00 &  346.00 &    74.50 &    76.00 &  312.153 &  363.027 &   6.807 &  14.518 &  $-$6.705 &  $-$2.860 \\
\hline
\end{tabular}
\end{center}
\end{table*}

\subsection{Pre-main-sequence stars not associated with molecular clouds}\label{sect:comoving}

After selecting YSO candidates associated with molecular clouds we analyzed the remaining pre-main-sequence star candidates from the catalogue of \citet{Zari2018}. Two large groups of stars at 140\,--\,200\,pc to the Sun show up in the lower panel of Fig.~\ref{fig1}. One of them at the eastern part of our studied region is projected onto the region devoid of molecular gas, bordering the Cepheus flare. Several publications have already studied comoving stars in this region. \citet{Guillout2010}  identified four lithium-rich comoving stars, and \citet{Faherty2018} listed seven stars in this region, three of which are identical with those in \citet{Guillout2010}. \citet{Klutsch2020} identified 29 further pre-main-sequence members of this group via high-resolution spectroscopy of X-ray source counterparts, and derived a mean distance of $157\pm10$\,pc and age of 10--20 million years. We adopt the name \textit{Cepheus Association\/} from their work. We note that while the gas-free region itself is thought to be a supernova bubble, interacting with the molecular clouds of the Cepheus flare, the young stars projected within this area are much closer to the Sun. The other group is located at the northwestern part of the Cepheus flare. We defined two rectangles (see Table\, \ref{tab:uvw} for the boundaries)
%(with $100\degr \leq l \leq 108\fdg2; 17\degr \leq b \leq 22\degr$ and $112\fdg5 \leq l \leq 124\degr; 11\degr \leq b \leq 21\fdg5$, see Figure \ref{fig1})
to search for additional group members. 

\begin{figure}
\includegraphics[width=\columnwidth]{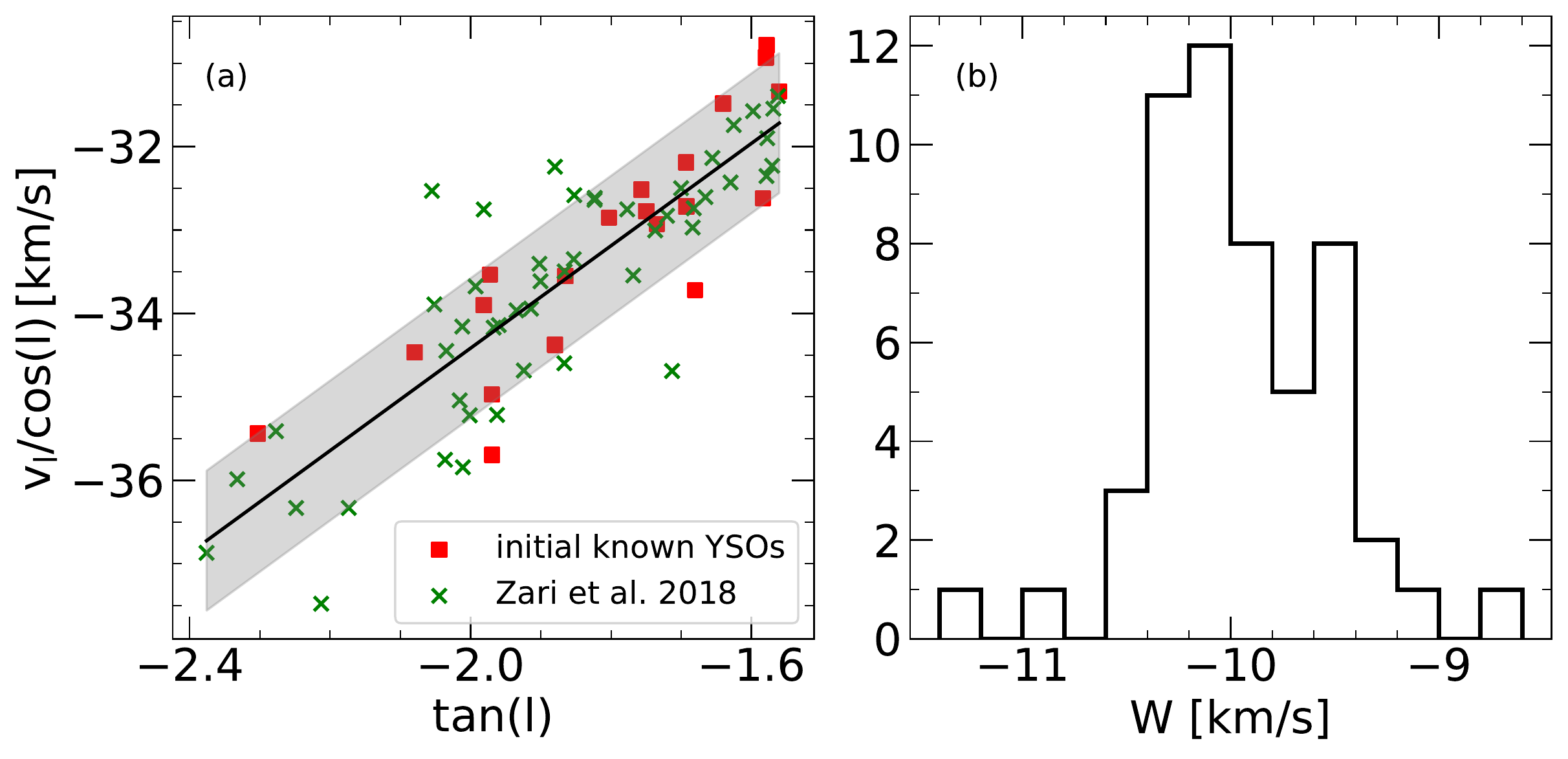}
\caption{\textsl{a}: Distribution of the members and new candidate members of the Cepheus Association in the ($tan(l)$,$v_\mathrm{l}/cos(l)$) plane.\textsl{b}: Histogram of the \textit{W\/} velocities computed for the stars in the shaded area of panel (a).} 
\label{fig:cepass_tanl}
\end{figure}

Since these groups stretch over several degrees, the proper motions of stars, having common space motions, may not be identical over the whole area, thus the method described in Section \ref{sel_crit} cannot be applied. 
To identify comoving stars within these large areas we followed the method described in \citet{Damiani2019}: stars having the same space motion  should fall along a straight line in the ($tan(l)$,$v_\mathrm{l}/cos(l)$) plane, whose slope and intercept provide the space velocity components \textit{U} and \textit{V} of the kinematic group, respectively. Figure~\ref{fig:cepass_tanl}a shows the distribution of the members of the Cepheus Association \citep[red squares, from][]{Klutsch2020}, and the candidate pre-main-sequence stars with $140\,\mathrm{pc} < d < 200\,\mathrm{pc}$ from \citet[][green crosses]{Zari2018} in the $v_\mathrm{l}/cos(l)$ vs. $tan(l)$ plot. The straight line is fitted to all points displayed. 
The grey band indicates the formal uncertainty of the fitted line. We selected the stars inside the grey bands, and using the \textit{U} and \textit{V} velocity components derived from the fitted line calculated their \textit{W\/} velocity component \citep[eq. 17 in][]{Damiani2019}. The new candidate members of the kinematic group were selected using the histograms of \textit{W\/}, displayed in Fig.~\ref{fig:cepass_tanl}b. The derived \textit{W\/} velocities are distributed over a narrow range mainly between $-$10.5 and $-$9\,km\,s$^{-1}$, thus we adopted all of them as candidate members. 

\begin{figure}
\includegraphics[width=\columnwidth]{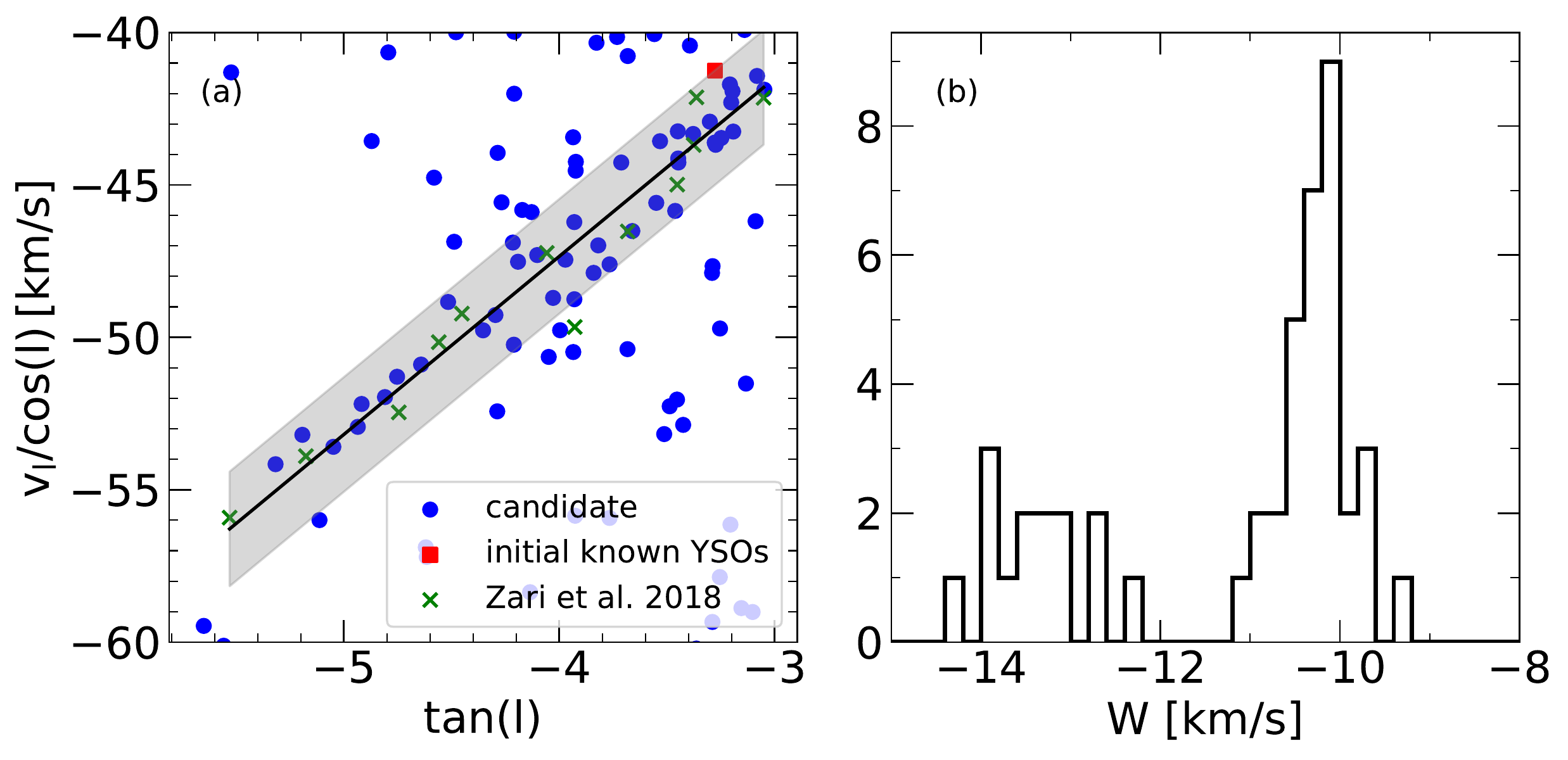}
\caption{Same as Fig.~\ref{fig:cepass_tanl}, but for the HD\,190833 moving group. }
\label{fig:hdgroup_tanl}
\end{figure}

We used the same method for selecting members from the catalogue of \citet{Zari2018} and \gaia~EDR3 between 160 and 200\,pc in the other area, labeled as \textit{HD\,190833 group\/} after the brightest candidate member. Figure~\ref{fig:hdgroup_tanl}a shows \gaia~sources with $v_\mathrm{l}/cos(l)$ between $-$60 and $-$40 km\,s$^{-1}$, where the candidate YSOs of \citet{Zari2018} are located. The straight line in Fig.~\ref{fig:hdgroup_tanl}a is fitted to the sources of \citet{Zari2018} and [TNK2005]\,20. Figure~\ref{fig:hdgroup_tanl}b shows that the \textit{W\/} distribution has two peaks at around $-$13.5 and $-$10 km\,s$^{-1}$, suggesting that the \gaia~DR3 stars, grouped around the fitted line of Fig.~\ref{fig:hdgroup_tanl}a, form two kinematic subgroups, which differ only in the velocity component \textit{W\/}, perpendicular to the Galactic plane.  We adopt the stars of both subgroups as candidate members of the HD\,190833 moving group. %Their surface and tangential velocity distribution are shown in Fig.~\ref{fig:hdkingroup}.

Figure~\ref{fig:Cepheus_Association}a and \ref{fig:hdgroup}a show the surface distributions and tangential velocities of of the candidate members of these nearby moving groups. The mean space velocity components of the two groups, derived during this process, are listed in Table~\ref{tab:uvw}.  \gaia~EDR3 identifiers, coordinates, distances, tangential velocities along Galactic longitude and latitude, and other names of the new candidate group members are listed in Table\,\ref{tab:yso-cand}.

\begin{table}
\begin{center}
\caption{Boundaries and mean \textit{U\/},  \textit{V\/}, \textit{W\/} velocities derived for the Cepheus Association and HD\,190833 group.}
\begin{tabular}{lcc}
\hline
& Cepheus Association & HD\,190833 group\\
\hline
${l_\mathrm{min}}$ & 112\fdg5 & 100\degr \\
${l_\mathrm{max}}$ & 124\degr & 108\degr \\
${b_\mathrm{min}}$ & 11\degr & 17\degr \\
${b_\mathrm{max}}$ & 21\degr & 22\degr \\
\textit{U\/} (\kms)& ~$6.1\pm0.4$ & ~~$5.8\pm0.5$\\
\textit{V\/} (\kms)& $-22.1\pm0.8$ & $-24.0\pm1.9$\\
\textit{W\/} (\kms) & $-10.0\pm0.4$ & $-11.2\pm1.5$\\
\hline
\end{tabular}
\label{tab:uvw}
\end{center}
\end{table}

\subsection{Characterising the YSO groups and new candidate members}
\label{sect3.3}
\paragraph*{Minimum spanning trees.}
We constructed minimum spanning trees (MST) to characterise the size and structure of the identified clusters and aggregates. Figures~\ref{fig:ngc7023}b--\ref{fig:hdgroup}b show the MSTs for each cloud.%\footnote{For the MST-s we used the following package: \url{https://github.com/jakevdp/mst\_clustering}} 

\begin{figure}
\centering \includegraphics[width=\columnwidth]{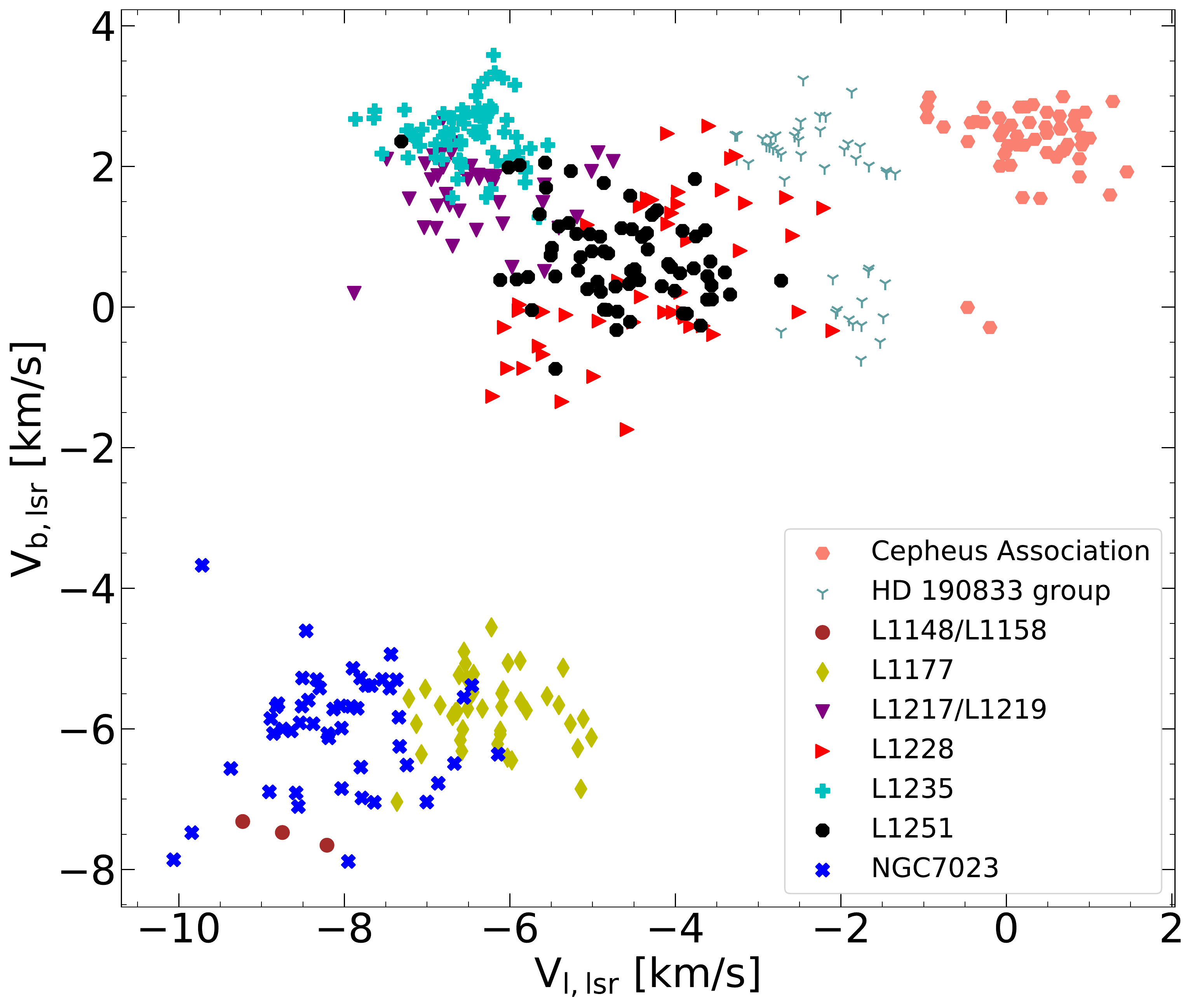}
\caption{Tangential velocities of the members and candidate members of the subregions of the Cepheus flare as compared to the Local Standard of the Rest. }
\label{fig:vel}
\end{figure}

\paragraph*{Tangential velocities.}
We calculated velocities compared to the \textit{Local Standard of Rest\/} (LSR) by subtracting from the tangential velocities the (\textit{l,b}) components of the Solar peculiar motion, using the  $U,V,W$ velocity components of Solar motion determined by \citet{Schonrich2010}.
Figure~\ref{fig:vel} shows the distribution of the members of individual groups in the $v_{b,\mathrm{LSR}}$ vs. $v_{l,\mathrm{LSR}}$ plane. The same distributions are displayed for each group separately in Figs.~\ref{fig:ngc7023}d--\ref{fig:hdgroup}d. The average velocity components of the whole system of candidate pre-main-sequence stars, corrected for the reflex solar motion, are \textit{v}$_{l,\mathrm{avg}}^\mathrm{LSR } = 8.3\pm2.4$\,km\,s$^{-1}$ and \textit{v}$_{b,\mathrm{avg}}^\mathrm{LSR} = -6.7\pm3.9$\,km\,s$^{-1}$.

\paragraph*{Colour--magnitude diagrams of the young clusters.}
To have an insight into the relative ages of the YSO groups, \textit{G} vs. (\textit{G}$-$\textit{G}$_{\mathrm{RP}}$) colour--magnitude diagrams of confirmed and candidate pre-main-sequence members of individual dark clouds are plotted in Figs.~\ref{fig:ngc7023}e--\ref{fig:hdgroup}e. Isochrones of 1, 10, and 20~million years of the PARSEC \citep{Bressan2012} models for masses above 1.4\,\msun\  and CIFIST \citep{BHAC2015} models for masses below 1.4\,\msun\ are also plotted. 

\paragraph*{2MASS colour--colour diagrams.}
To assess the nature of the newly identified candidate pre-main-sequence stars we supplemented the \gaia\ data with 2MASS \citep{2mass} data. To search for 2MASS counterparts of the stars, we transformed ICRS coordinates from \gaia's J2016 epoch into J2000 epoch. Then we searched for coinciding 2MASS sources within 1\,arcsec. Several binary systems in the initial list have separations larger than 1\,arcsec, but were identified as single 2MASS sources. In those cases we manually crossmatched the 2MASS with the \gaia\ data. \jb$-$\h\ vs. \h$-$\ks\  colour-colour diagrams of the clusters are shown in Figs.~\ref{fig:ngc7023}f--\ref{fig:hdgroup}f.

\begin{figure}
\centering \includegraphics[width=\columnwidth]{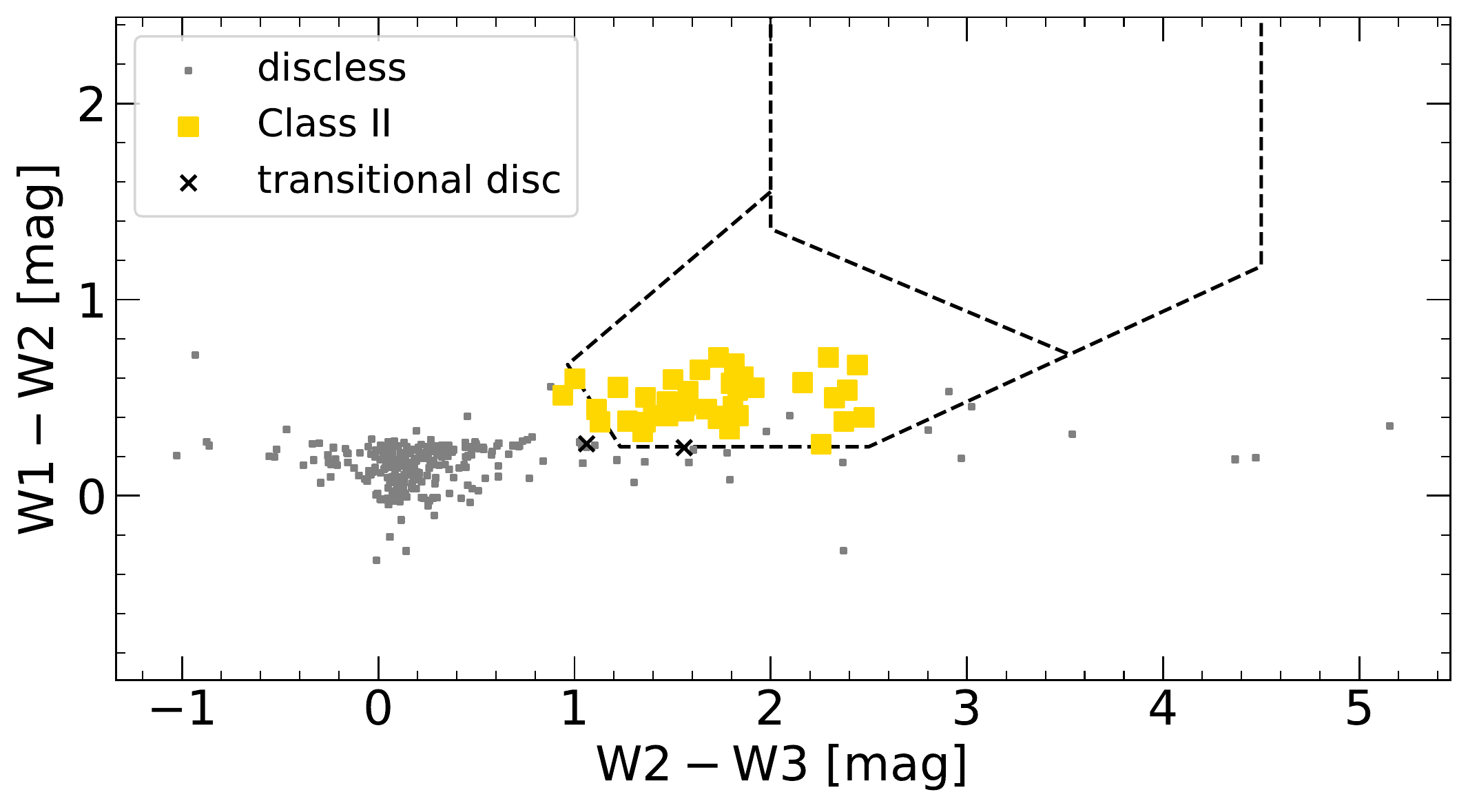}
\caption{\textit{AllWISE\/} colour--colour diagram of the newly identified candidate YSOs of the Cepheus flare. Dashed lines border the areas occupied by Class~I and Class~II YSOs, according to \citet{Koenig2014}. }
\label{fig:wise}
\end{figure}

\paragraph*{\wise\ colour--colour diagram.}
Mid-infrared colour indices are helpful in identifying disc-bearing pre-main-sequnce stars. Whereas \spitzer\ observed only a small fraction of the studied area, \wise\ \citep{Wright2010} data at 3.4, 4.6, 12.0, and 22.0 \micron\ are available for the whole region in the \textit{AllWISE\/} data base.
We found \wise\ counterparts of 313 candidate YSOs within 1\,arcsec. To identify infrared-excess stars, we followed the methods described by \citet{Koenig2014}. Of the 313 \wise\ sources, 225 fulfil the quality criteria essential for constructing their ($w1-w2$) vs. ($w2-w3$) colour-colour diagram, displayed in Fig.~\ref{fig:wise}. Thirty-five of the displayed points lie within the area of the Class~II sources, defined by \citet{Koenig2014}.
Involving 2MASS data the ($H-K_\mathrm{s}$) vs. ($w1-w2$) diagram resulted in three further Class~II sources, and two candidate transitional discs were identified in the ($w1-w2$) vs. ($w3-w4$) diagram. The candidate disc-bearing YSOs are listed in Table~\ref{tab:allwise}.  We encircled these sources with black diamonds in Figs.~\ref{fig:ngc7023}a--\ref{fig:hdgroup}a.   

\paragraph*{Interstellar extinction towards the YSO groups.}
We obtained reddening versus distance curves for the nominal \textit{Simbad} positions of the molecular clouds from the 3D maps of \citet{green2019}, except for HD\,190833 group, where we used the central coordinates of the tetragon. We used the Python-implemented
{\fontfamily{pcr}\selectfont
dustmap} package with the best fitted model to download the reddenings and $R_V=3.1$ to transform them into \av\ extinctions (see Figures\,\ref{fig:ngc7023}a--\ref{fig:hdgroup}g).

\section{Results for the individual molecular clouds and YSO groups}
\label{sect4}

\begin{figure*}
\includegraphics[width=14cm]{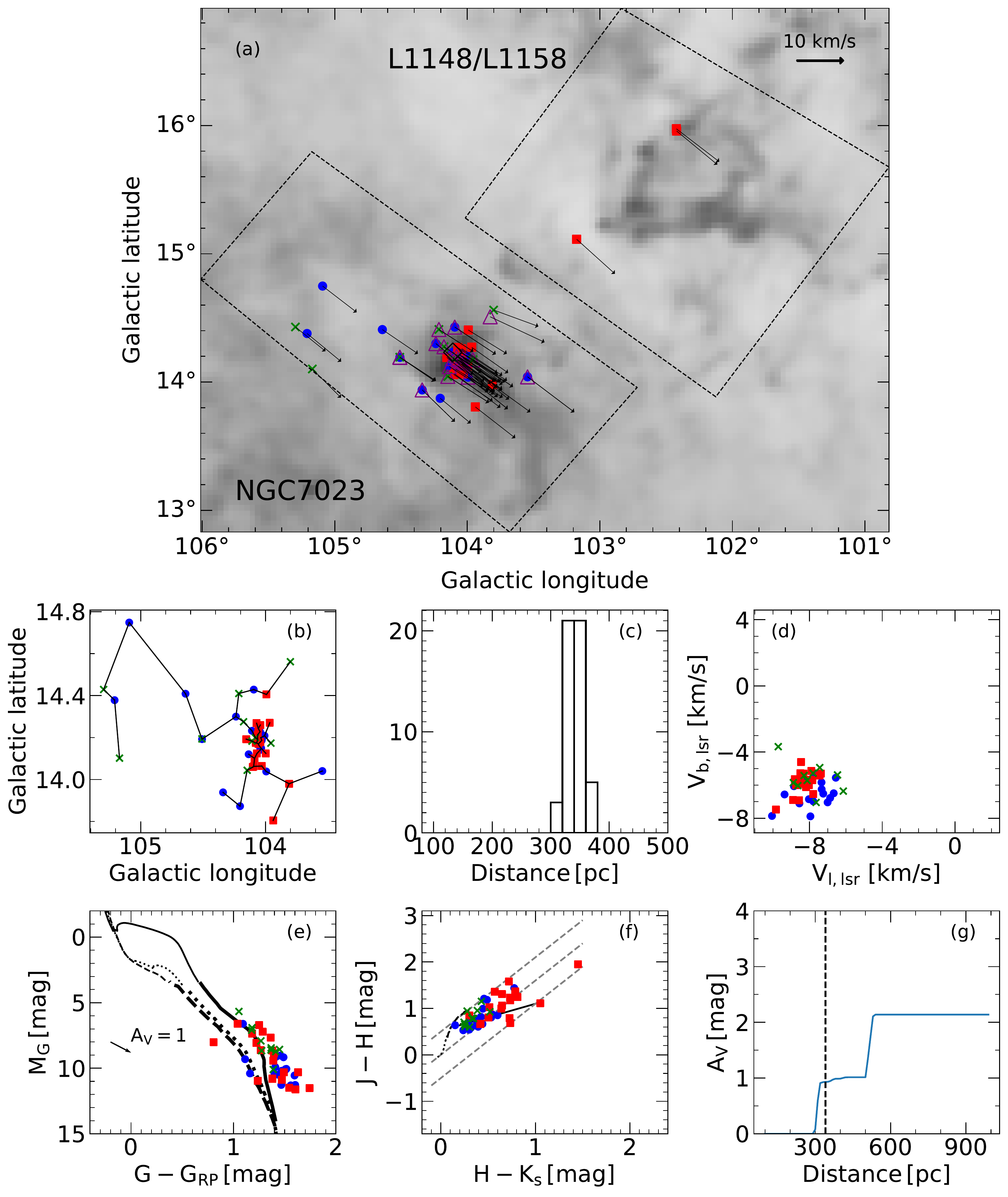}
\caption{\textsl{a}: Distribution of known YSOs (red squares), YSO candidates (blue circles) and candidate pre-main-sequence stars from \citet{Zari2018} (green crosses) in NGC\,7023 and in L1148/L1158, plotted on the \textit{Planck\/} 857\,GHz image. The arrows indicate the tangential velocities of the stars compared to the LSR. Colour codes are same for all other subplots. The black diamond around the filled circle indicates the candidate YSO exhibiting infrared excess in the \wise\ bands. The purple triangles represent candidate cluster members identified by \citet{Saha2020}. \textsl{b}: Minimal spanning tree of the stars plotted in panel \textsl{b}. \textsl{c}: Histograms of distances. \textsl{d}: Galactic longitudal vs. latitudal components of tangential velocities compared to the LSR. \textsl{e}: $M_\mathrm{G}$ vs. $G-G_\mathrm{RP}$ colour--magnitude diagram. The thick and thin lines are \textit{CIFIST\/} and \textit{PARSEC\/} isochrones, respectively. The solid, dotted, and dashed lines are isochrones of 1, 10, and 20 million years, respectively. \textsl{f}: Colour--Colour-diagram of stars with 2MASS identifiers. \textsl{g}: Visual extinction \textit{A\/}$_\mathrm{V}$ at the field centre as a function of the distance, obtained from \citet{green2019}. The dashed line shows the mean distance of the stars.}
\label{fig:ngc7023}
\end{figure*}

\paragraph*{L1148/L1158.} Our initial list of YSOs contains 11 objects projected on this cloud complex. Eight of them fulfil our astrometric quality criteria. Only three of the eight -- RNO\,124, SSTgbs~J2048103+6803019, and SSTgbs~J2036116+6757093 -- have distances compatible with the cloud complex. (The fourth known optically visible YSO of the region, PV~Cep has $\mathrm{RUWE} = 2.4$, indicative of uncertain astrometric data.) Five of the 8 sources have distances between 607 and 1080\,pc. All of them were detected only by the MIPS instrument of \spitzer\ at 24\,\micron, and were classified as off-cloud objects by  \citet{Kirk2009}. They suggest the presence of a star-forming region between these distances. The \av\ vs. \textit{D} diagram in Fig.~\ref{fig:ngc7023}g and the catalogue of \citet{Zucker2020} also indicate the presence of a distant layer of obscuring matter towards some nearby lines of sight. 

The 12 candidate PMS stars in \citet{Zari2018}, projected onto these clouds, differ in parallax and proper motion from the confirmed YSOs and from each other. The \textit{Planck} 857\,GHz map of the L1148/L1158 cloud complex and its associated \gaia\ stars are shown in Fig.~\ref{fig:ngc7023}a, together with the neighbouring NGC\,7023.

\paragraph*{NGC 7023.} This region was recently studied in detail by \citet{Saha2020}. They identified 20 new members of the cluster based on \gaia~DR2 data, and derived a distance of $335\pm11$\,pc. Our initial list contain 37 spectroscopically or \spitzer-classified YSOs, projected within the tetragon containing the NGC\,7023 cluster, 22 of which fulfil our astrometric quality criteria. One of them, SSTgbs~J2104156+6742464 is located far beyond the cluster. When defining the tangential velocity ranges, we used 19 stars within the intervals listed in Table~\ref{tab:filter}. These stars are plotted with red symbols in Fig.~\ref{fig:ngc7023}).  Based on \gaia~EDR3 data we identified 31 comoving stars within the defined tetragon (green and blue symbols in Fig.~\ref{fig:ngc7023}). Ten of the 31 are found in the catalogue of \citet{Zari2018}. The mean distance of the cluster +/- the standard error of the mean is $341\pm2$\,pc. The colour--magnitude diagram (Fig.~\ref{fig:ngc7023}e) suggest moderately reddened low-mass YSOs. Taking into account an average foreground extinction of 1.6\,mag \citep{Saha2020} we estimate an age of some 1~million years. The star below the 20-Myr isochrone of the colour--magnitude diagram in Fig.~\ref{fig:ngc7023}e is Cl*NGC\,7023\,RS\,10, whose spectral energy distribution suggests a Class\,I YSO, and thus its ($G-G_\mathrm{RP}$) colour index is affected by scattered light from the disc atmosphere. Fig.~\ref{fig:ngc7023}f shows that the new, comoving members are less reddened than the initial members. Neither of them has \ks-band excess, nor \wise\ colour indices characteristic of YSOs. They may represent the discless population of the cluster. We estimated the fraction of disc-bearing stars so that spectroscopically confirmed YSOs with insufficient \gaia~EDR3 data were included and the comoving stars outside of the cluster were discarded. Thus we obtained N$_\mathrm{disc}$/N$_\mathrm{total}= 0.60$.

\begin{figure*}
\includegraphics[width=14cm]{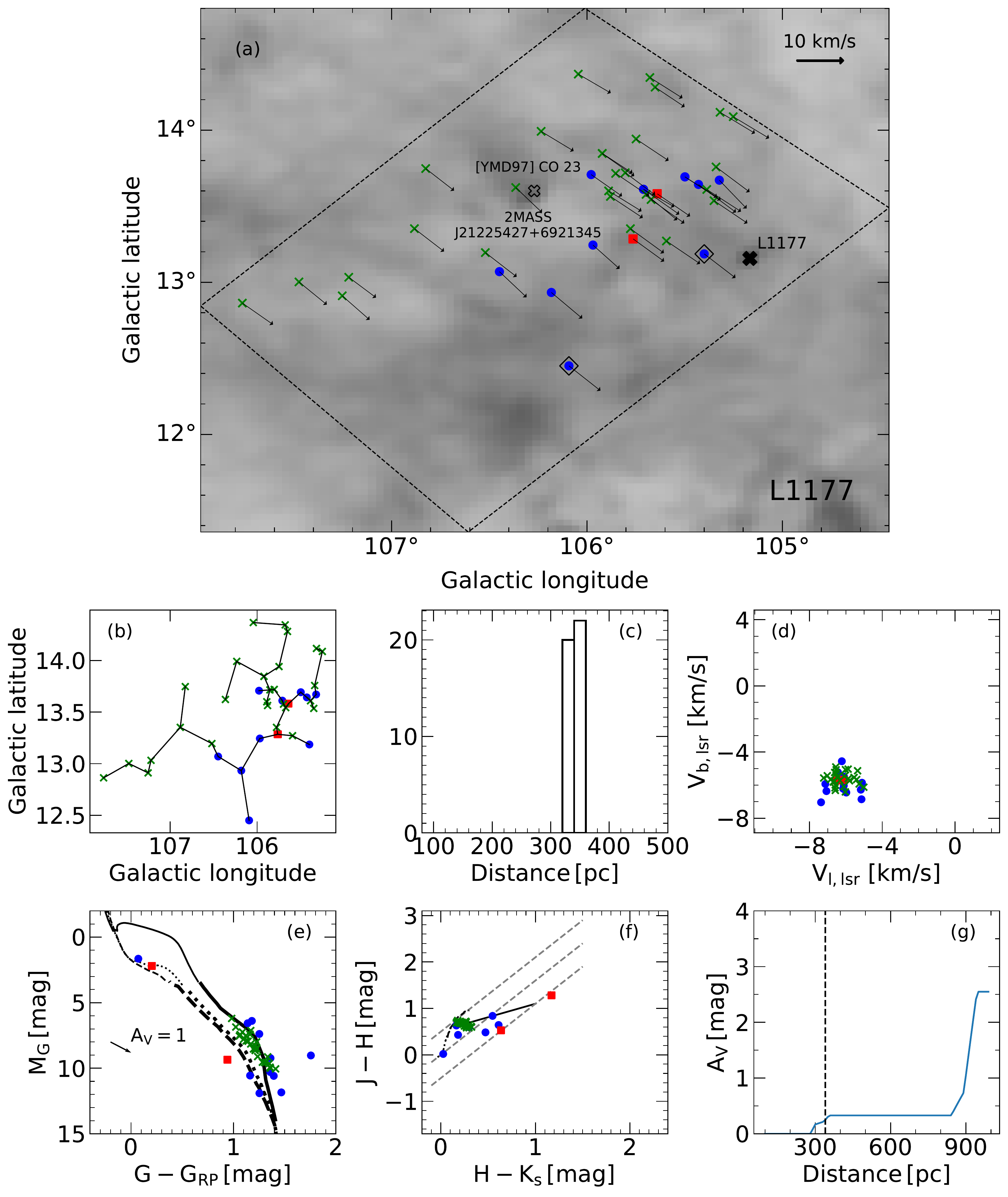}
\caption{Same as Figure \ref{fig:ngc7023} for L1177.}
\label{fig:l1177}
\end{figure*}

\paragraph*{L1177.} This small globule at (\textit{l,b})=(105\fdg17,+13\fdg16) contains a single YSO, the FUor-like star 2MASS~J21173862+6817340 \citep{Connelley2018}. The lower panel of Fig.~\ref{fig1} suggests a condensation of candidate pre-main-sequence stars, catalogued by \citet{Zari2018}, to the north of the globule.
We defined a rectangle to encompass the globule and this apparent group of candidate PMS stars. Three confirmed PMS stars of the cloud complex (BD +68\degr1118, [K98c]\,Em*53, and [K98c]\,Em*58) are found within the rectangle. The distance and tangential velocity intervals were defined by BD +68\degr1118, [K98c]\,Em*58 and 28 stars from \citet{Zari2018}, and we found 12 further candidate members. [K98c]\,Em*53 was excluded due to the large  errors of its proper motion and RUWE.
The most luminous members of the group (Fig.~\ref{fig:l1177}e) are the A0-type HD\,203533 and the Herbig~Ae star BD+68\degr1118. Figure~\ref{fig:l1177}e suggests a mean age of some 5~million years. The star below the 20-Myr isochrone is the Class\,I YSO [K98c]\,Em*58. All these comoving stars are found in a low-extinction region, and according to Fig.~\ref{fig:l1177}f, all but two of them are M-type stars without \ks-band excess. The \ks-band-excess stars, 2MASS~21190541+6828465 and 2MASS J21295798+6827007 are new candidate classical T~Tauri stars. The colour indices of their \wise\ counterparts confirms its CTTS nature.  Figure~\ref{fig:l1177}g shows that most of the extinction in this region originates from the distant cloud (the high-velocity sheet in \citet{Heiles1967} and \citet{Grenier1989}). Figure~\ref{fig:l1177}a demonstrates that distant star-forming clouds, belonging to the high-velocity complex around 900\,pc, are present among the nearby clouds. The molecular cloud [YDM97]\,CO\,23 \citep{YDM1997}, marked in Fig.~\ref{fig:l1177}a, with its radial velocity of $v_\mathrm{LSR}=-9.7$\,km\,s$^{-1}$ is part of the distant cloud layer, and the birthplace of the distant classical T~Tauri star 2MASS~J21225427+6921345 \citep{kun2009} ($\varpi = 0.961\pm0.110$\,mas). 

\begin{figure*}
\includegraphics[width=14cm]{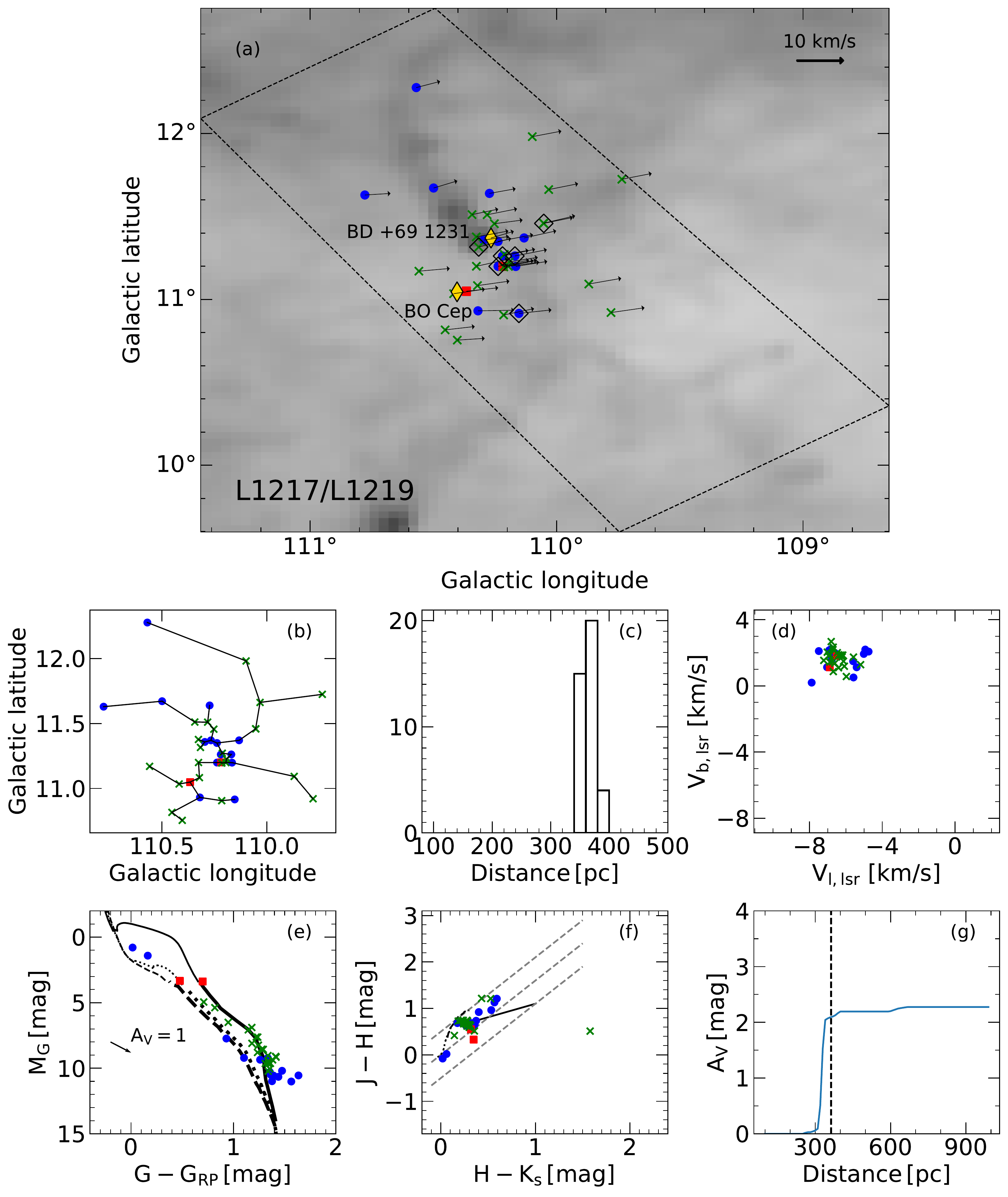}
\caption{Same as Fig.~\ref{fig:ngc7023} for L1217/1219.}
\label{fig:l1217}
\end{figure*}

\paragraph*{L1217/L1219.} L1219 (Barnard 175) has not been observed by \textit{Spitzer\/}, therefore its YSO population is poorly known. Three T~Tauri stars listed in \citet{kun2009} and the Herbig~Fe star BO~Cep are projected near the cloud. One of the four YSOs (TYC 4467-324-1) does not fulfil our quality criteria, and another one (BO~Cep) differs in the direction of motion from the others. We defined the intervals in Table~\ref{tab:filter} based on the data of two known T~Tauri stars and 23 candidate PMS stars from \citet{Zari2018}, and identified 16 further comoving stars clustered around L1219. BD+69\degr1231, the illuminating star of the bright reflection nebula LBN~110.25+11.38 (Ced 201), and HD\,211772, both classified by \citet{Zari2018} as young upper main sequence stars, are members of the group. The average distance of the cluster is $363\pm2$\,pc. Figure~\ref{fig:l1217}f shows that most of the new candidate members are unreddened M-type stars without \ks-band excess. Six of them have excess emission, characteristic of Class\,II YSOs, in the \wise\ photometric bands. Figure~\ref{fig:l1217}e suggests an age of 1--5~million years.

\begin{figure*}
\includegraphics[width=14cm]{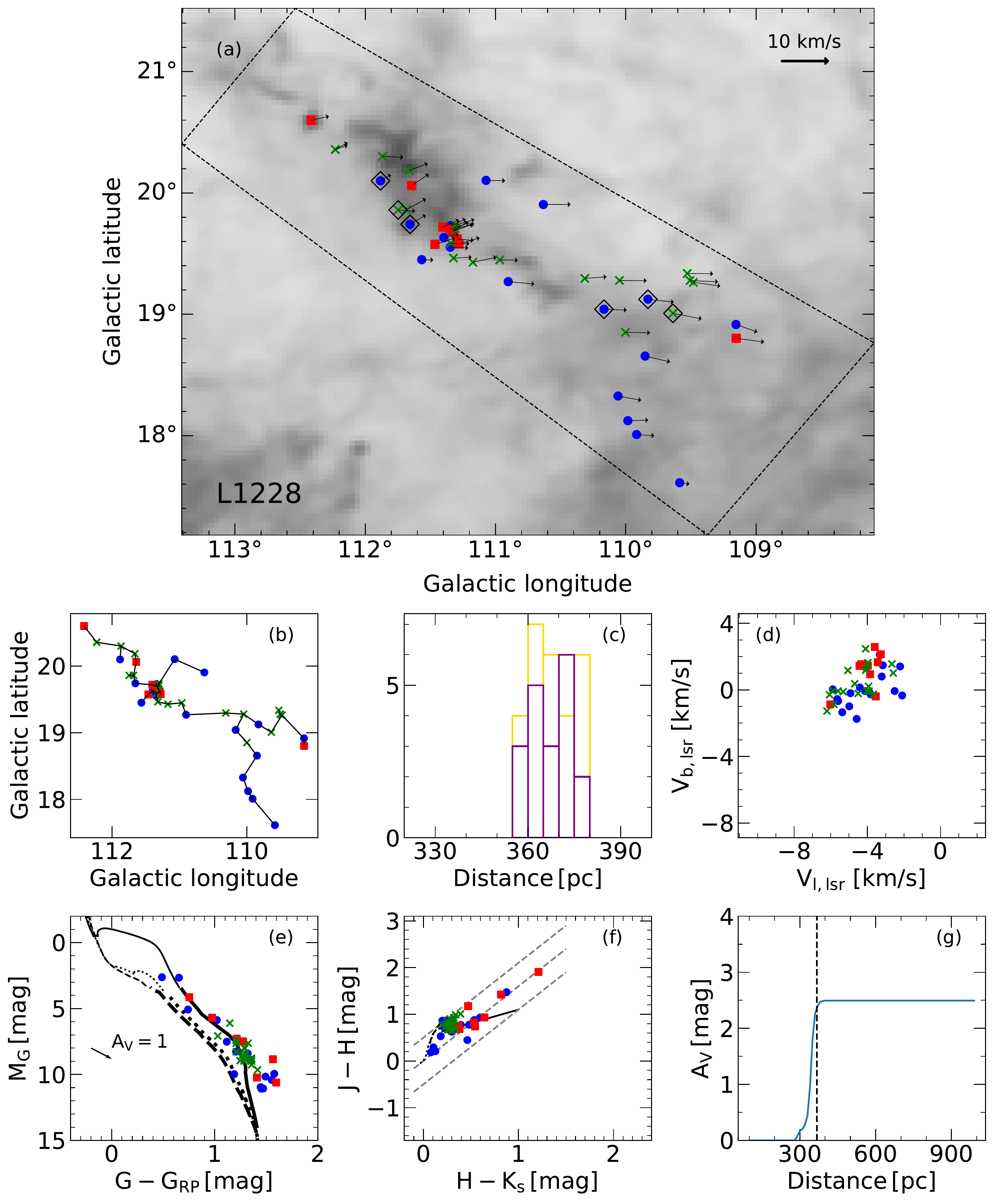}
\caption{Same as Fig.~\ref{fig:ngc7023} for L1228. The purple and gold colors on panel \textsl{c} show the distances of northern and southern subgroups, respectively.}
\label{fig:l1228}
\end{figure*}

\paragraph*{L1228.} L1228 is the northernmost cloud of the Cepheus Flare, and has the largest negative radial velocity \citep[$-$7.6\,\kms,][]{YDM1997}. The clustering of young stars and candidates projected around L1228 apparently extends beyond the boundaries of the molecular cloud, towards the south-west, where the small, diffuse molecular cloud [YDM97]\,CO\,60 (110.5+19.2) and the CTTS 2MASS~J20530638+7450348 can be found.
We included this cloud and the stars around 2MASS~J20530638+7450348 into the tetragon. Our initial list contains 27 YSOs projected within this area, 17 of which fulfil our astrometric quality criteria. Two of the 17, %SSTgbs~J2057130+7735437 and SSTgbs~J2101395+7706165 are located beyond 500\,pc, whereas 
SSTgbs~J2100380+7706598 and [TNK2005]~28 are foreground stars at 291 and 218\,pc, respectively. The remaining 15 stars are distributed over a wide distance range between 330 and 404\,pc. These stars are probably related to the star-forming history of the region. However, the search for comoving cluster members based on the whole sample resulted in several irrelevant objects, therefore we defined narrower intervals (see Table~\ref{tab:filter}), taking into account only nine stars, projected onto the central ridge of the cloud  and  2MASS~J20530638+7450348, located at the south-eastern corner of the region.  These ten stars are plotted with red squares in Fig.~\ref{fig:l1228}. Their average distance of $368\pm1$\,pc renders L1228 the most distant cloud of the Cepheus flare.  Using the distance and velocity intervals defined in Table~\ref{tab:filter} we identified 36 new candidate members. We found that some stars, mainly in the southern part of the area (Fig.~\ref{fig:l1228}a), including GSC 04472-00143 differ in tangential velocities from those projected onto L1228 (L1228N), therefore we divided the stars into two kinematic groups with $v_{b,\mathrm{LSR}} = 0.5$\,\kms. The spatial distribution of both groups is displayed in  Fig.\,\ref{fig:l1228kingroup}, and their average tangential velocities and distances are shown in Table\,\ref{tab:subgroup_data}. Figure~\ref{fig:l1228kingroup} shows that the groups partly overlap, and Fig.\,\ref{fig:l1228}c shows that they have the same distance. These stars were probably born in the molecular cloud  [YDM97]\,CO\,60, which slightly differs from L1228 in radial velocity.
Figure~\ref{fig:l1228}e suggests an age of 1~million years. Figure~\ref{fig:l1228}f shows that most of the new candidate members are low-mass pre-main-sequence stars, but two stars earlier in spectral type than typical T~Tauri type stars are also apparent in these figures. One of them is  BD+76\degr825, associated with the reflection nebula GN~21.01.8, and the other is TYC 4590-843-1. Both stars are situated near the cloud centre and are coincide in proper motion with the known YSOs of the cloud. Six candidate group members, located outside of the area observed by \spitzer, have infrared excesses characteristic of Class~II YSOs in the \wise\ bands. The fraction of disc-bearing stars is about 52.6\% for the northern group, and 22\% for the southern group, suggesting the more evolved nature of the latter.  

\begin{table}
\begin{center}
\caption{Average distances and tangential velocities (compared to the LSR) of the two kinematic subgroups of L1228 and HD\,190833 group.}
\begin{tabular}{lccc}
\hline
Group & $\mathrm{d_{avg}}$ & $\mathrm{v_{l,lsr,avg}}$ & $\mathrm{v_{b,lsr,avg}}$\\
& (pc) & (km\,s$^{-1}$) & (km\,s$^{-1}$) \\
\hline
L1228 North & $366.93\pm6.47$ & $-3.66\pm0.72$ & ~~$1.55\pm0.48$\\
L1228 South & $367.98\pm6.92$ & $-4.71\pm1.09$ & $-0.38\pm0.51$\\
HD\,190833 & $184.65\pm8.01$ & $-1.84\pm0.33$ & $-0.05\pm0.39$\\
& $177.51\pm6.97$ & $-2.42\pm0.55$ & ~~$2.32\pm0.32$\\
\hline
\end{tabular}
\label{tab:subgroup_data}
\end{center}
\end{table}

\begin{figure*}
\includegraphics[width=14cm]{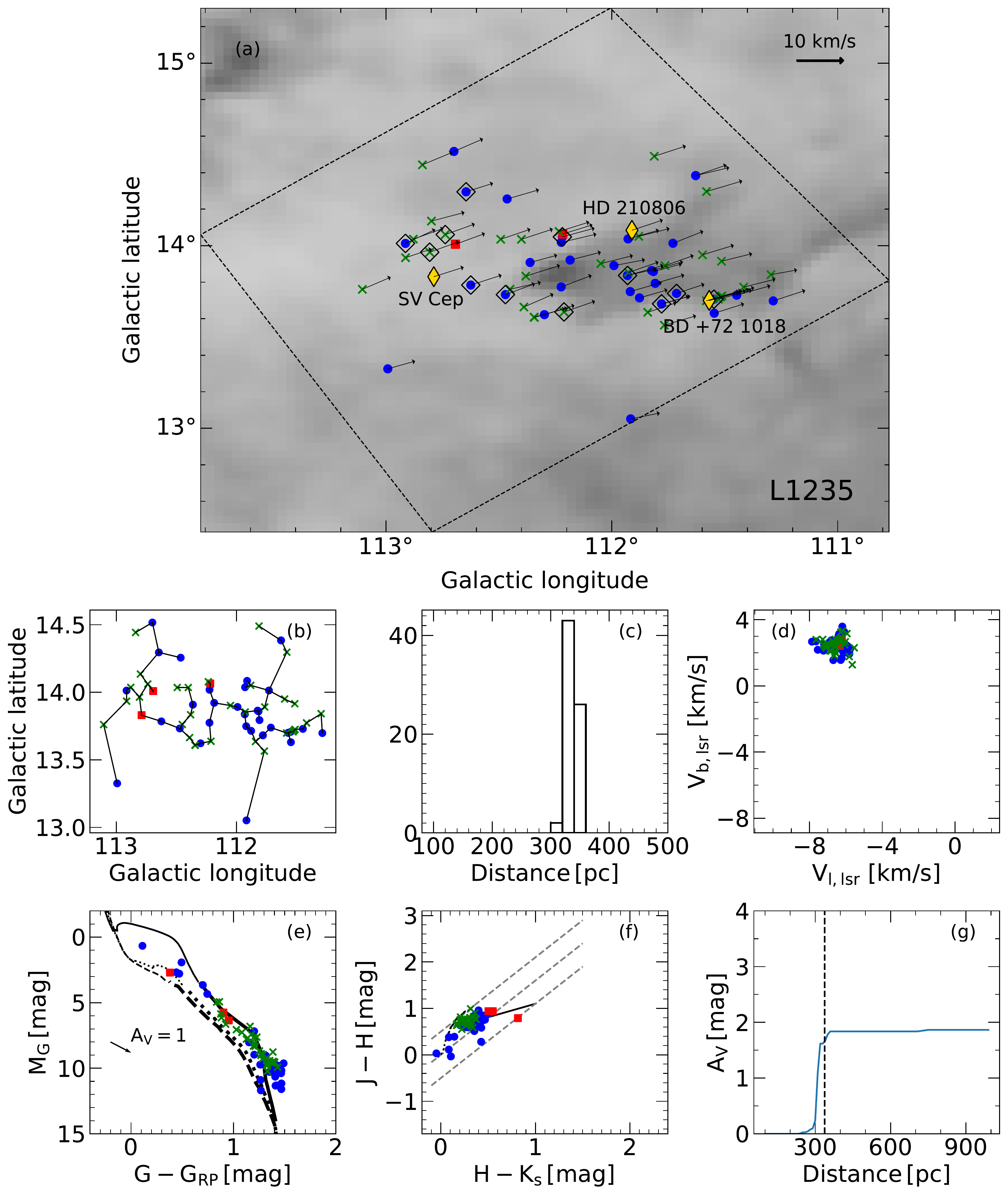}
\caption{Same as Fig.~\ref{fig:ngc7023} for L1235.}
\label{fig:l1235}
\end{figure*}

\paragraph*{L1235.}  L1235 has not been observed by \textit{Spitzer\/}, therefore its YSO population is less well mapped. The known young stellar population associated with L1235 consists of three classical T~Tauri stars, listed in  \citet{kun2009}, and the Herbig~Ae star SV~Cep. More recent star formation in the core of the cloud is indicated by the submillimeter sources observed by the SCUBA instrument on the JCMT \citep{scuba}. Our search revealed 68 candidate members, coinciding in parallax and proper motion with the T~Tauri stars listed in \citet{kun2009}, and also with the B8-type stars BD\,+72\degr1018 and HD\,210806,  illuminating reflection nebulae embedded in L1235. Thirty-two of the 68 candidate members are included in \citet{Zari2018}, and 12 of them have Class\,II-like infrared excess in the \wise\ bands, including the \gaia\ Photometric Science Alert source Gaia\,19bny. Figure~\ref{fig:l1235}e suggests ages 1--5~million years.

\begin{figure*}
\includegraphics[width=14cm]{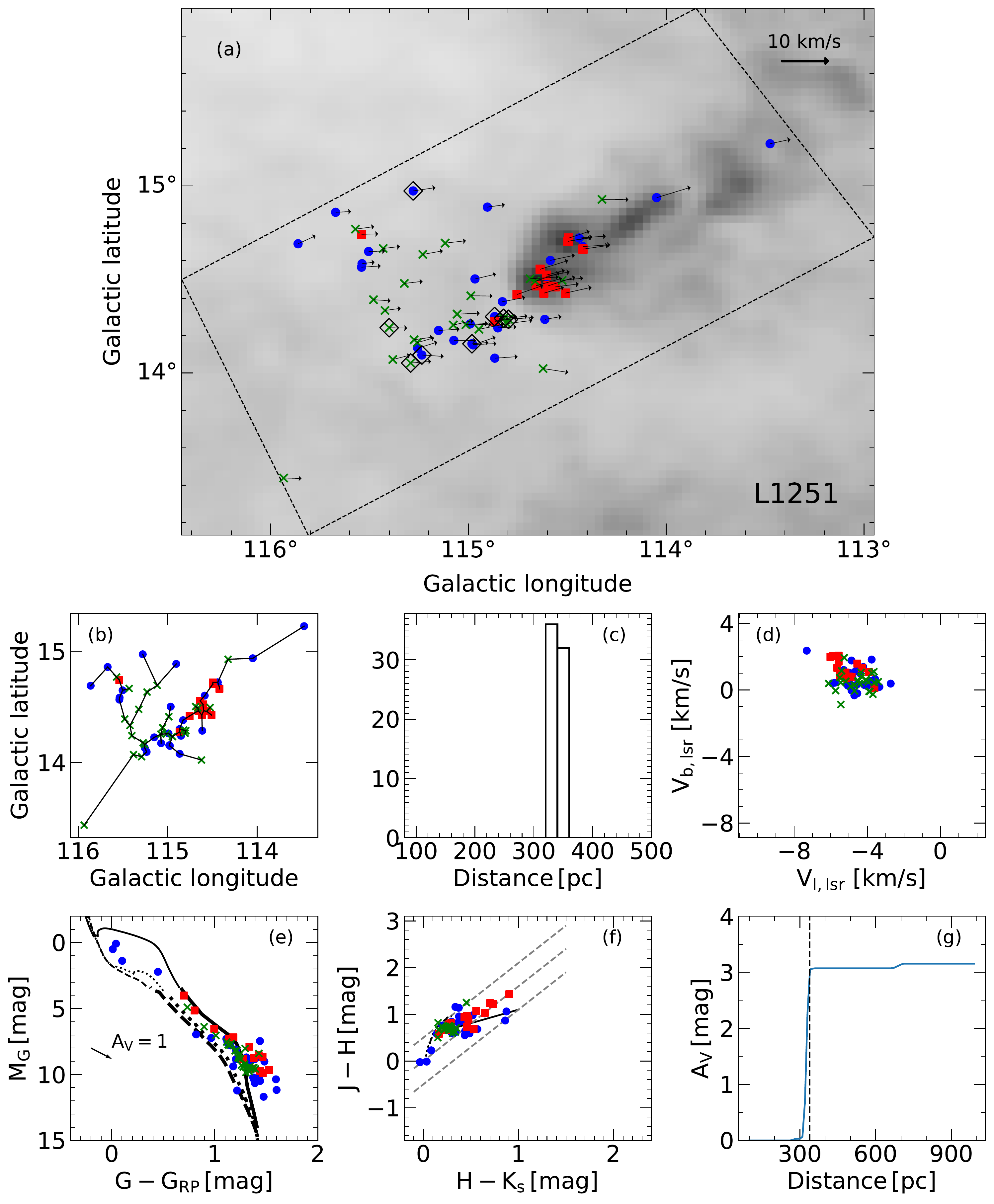}
\caption{Same as Fig.~\ref{fig:ngc7023} for L1251.}
\label{fig:l1251}
\end{figure*}

\paragraph*{L1251.} Our initial list of YSOs contains 31 stars associated with L1251. Fourteen of them fulfil our astrometric criteria. These stars are plotted with red squares in Fig.~\ref{fig:l1251}. We identified 54 stars coinciding in distance and velocity with these stars, and 26 of them is included in the catalogue of \cite{Zari2018}. Figure~\ref{fig:l1251}a shows that most of the new candidate members are located outside of the cloud. Figure~\ref{fig:l1251}e suggests an age of 1--5~million years, and \ref{fig:l1251}f shows that most of the new candidate members are M-type pre-main-sequence stars. Six of them exhibit infrared excess, characteristic of Class~II YSOs in the \wise\ wavelength region, and two of them bear transitional discs. The highest mass star of the group is the A0 type star HD\,216367, classified as young upper-main-sequence star by \citet{Zari2018}. 
The fraction of disc-bearing stars is apparently higher inside the cloud  (28/41) than outside (9/37), indicating the more evolved nature of the off-cloud population, but these numbers are affected by the different sensitivities of \spitzer\ and \wise. 

\subsection{Pre-main-sequence stars nearer than 200 pc}

\begin{figure*}
\includegraphics[width=14cm]{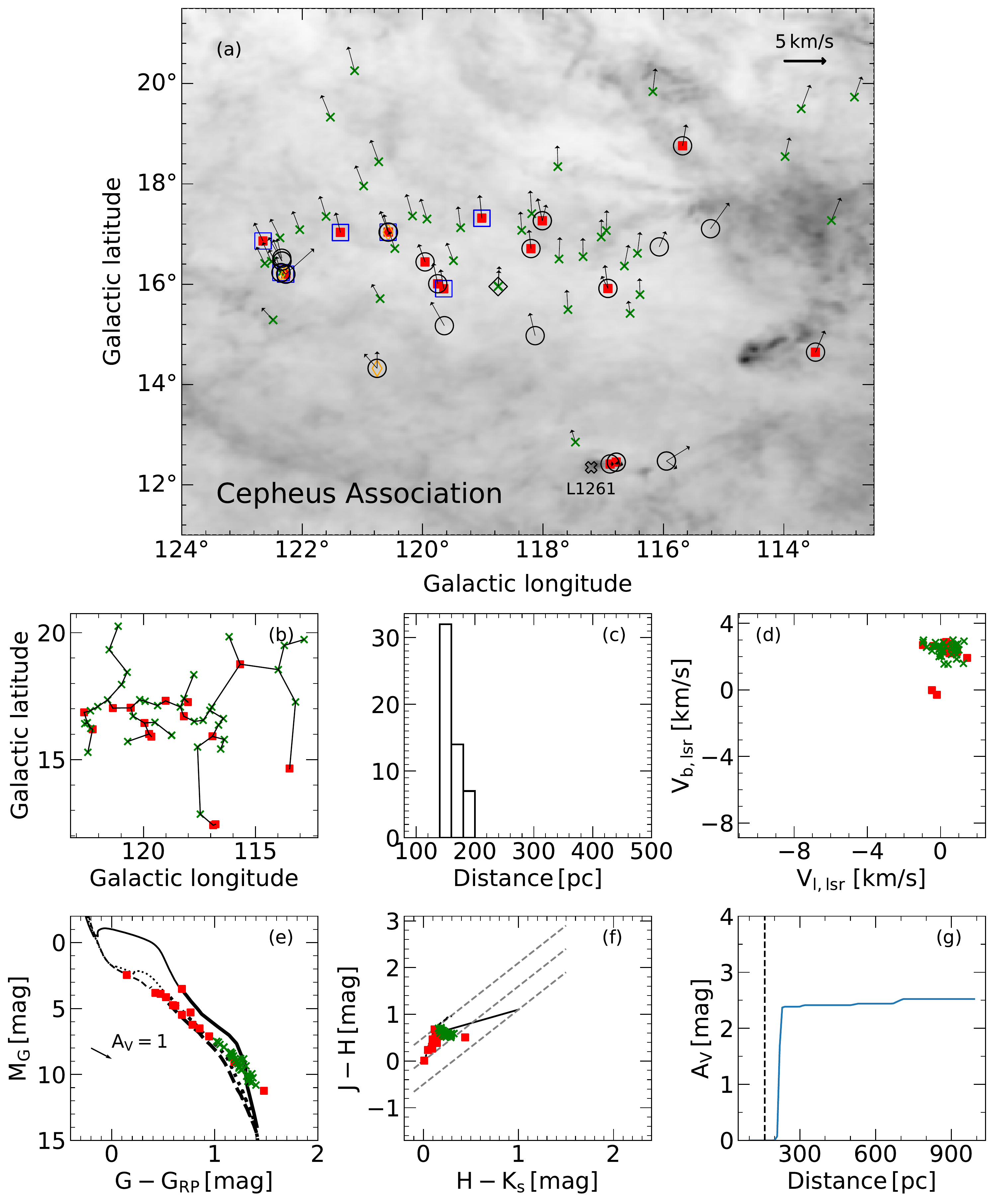}
\caption{Same as Fig.~\ref{fig:ngc7023} for the Cepheus association. The orange diamonds, blue squares and black empty circles are showing sources from \citet{Guillout2010}, \citet{Faherty2018} and Cepheus association members from \citet{Klutsch2020}, respectively.}
\label{fig:Cepheus_Association}
\end{figure*}

\begin{figure*}
\includegraphics[width=14cm]{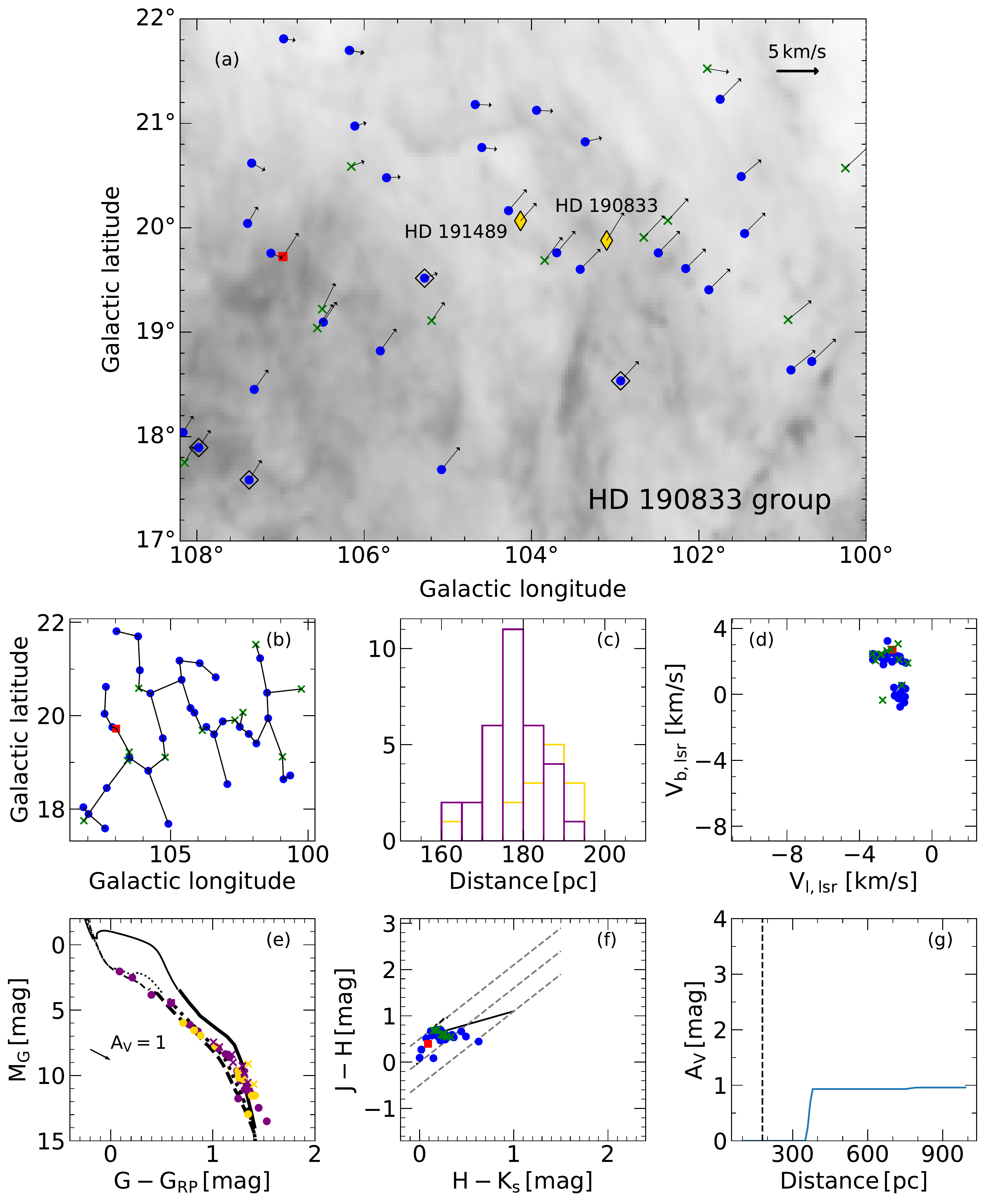}
\caption{Same as Fig.~\ref{fig:ngc7023} for the HD\,190833 group. The purple and gold colours on panel \textsl{c} and \textsl{e} show the distance histograms and the colour-magnitude diagrams of the two kinematic subgroups, respectively.}%, displayed in Fig.\,\ref{fig:hdkingroup}.}
\label{fig:hdgroup}
\end{figure*}

\begin{figure}
\includegraphics[width=\columnwidth]{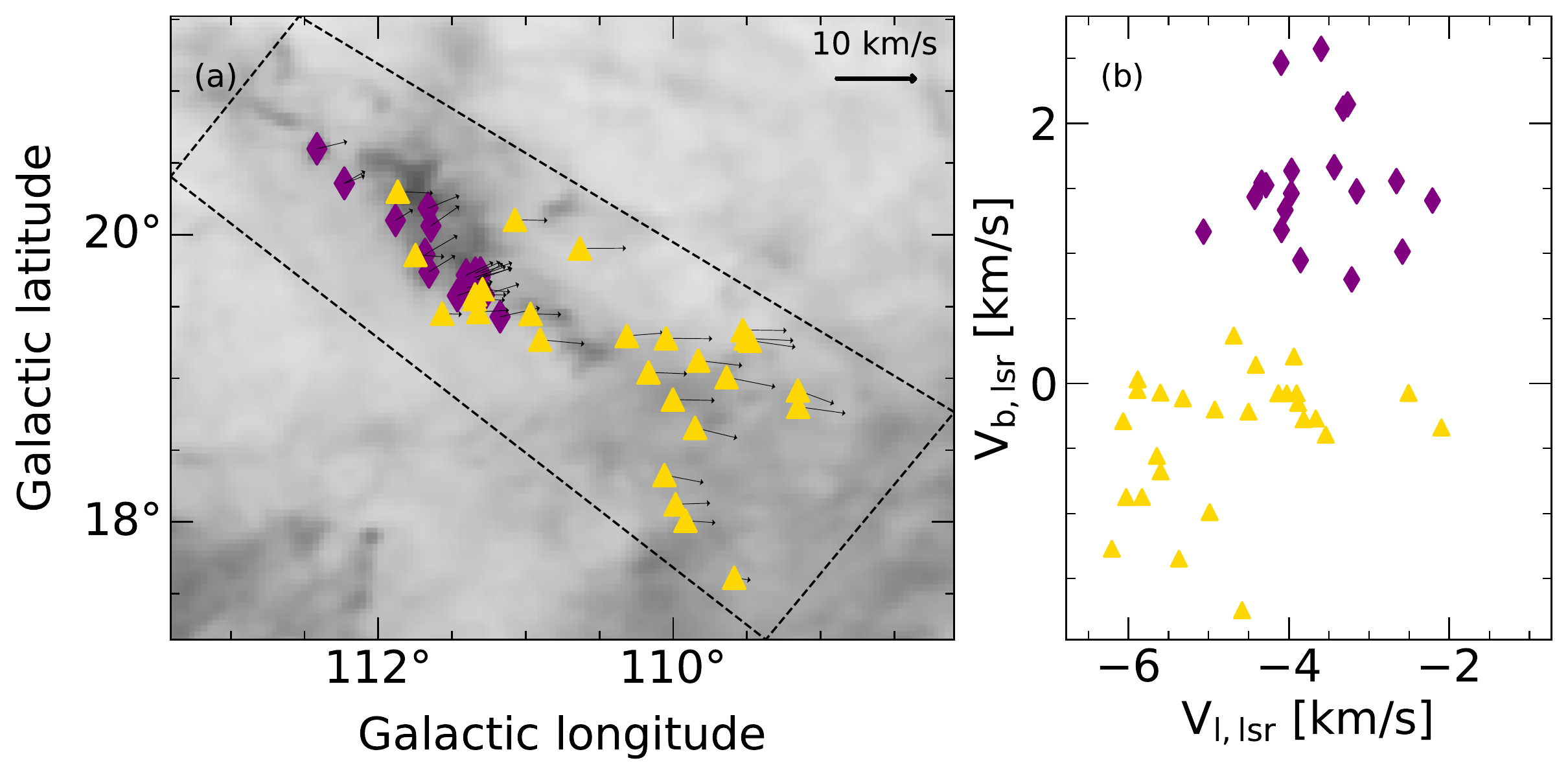}
\caption{Surface distribution and tangential velocities of two kinematic subgroups of L1228.}
\label{fig:l1228kingroup}
\end{figure}

\paragraph*{L1262 (CB 244).} This small globule contains the low-luminosity embedded protostar IRAS~23238+7401 \citep{Pollanen1995}.  Two optically visible YSOs, AS\,507 and [K98c]\,Em*137, projected on this cloud, have distances of 189 and 197\,pc, respectively, thus we adopt the distance 194\,pc for the cloud. AS\,507 has a companion at 9.3~arcsec separation \citep{kun2009}. The \gaia\ data revealed that both stars have similar distances and tangential velocities. L1262 is the only molecular cloud projected within the volume of the Cepheus Association.  However, only AS\,507 fulfils the kinematic membership criteria. 

\paragraph*{Cepheus Association.}
 We identified 37 new candidate members of the association within the $11\degr\times10\degr$ area indicated in Fig.~\ref{fig1} (see Sect.~\ref{sect:comoving}). These stars have a mean distance ($\pm$ standard error of the mean) of $161\pm2$\,pc. Figures~\ref{fig:Cepheus_Association}e and \ref{fig:Cepheus_Association}f show that the new candidate members represent the M-type population of the association. The two apparent outliers in Fig.~\ref{fig:Cepheus_Association}e correspond to the two stars located near the low-latitude boundary of the region, AS~507 and 2MASS~J23222636+7414115. According to Fig.~\ref{fig:hdgroup_tanl} their space velocities are consistent with those of the other association members. The divergent $v_\mathrm{b}$ positions result from their different Galactic latitudes. Fig.~\ref{fig:Cepheus_Association}e suggests an age of some 20~million years. These stars are older than the stars of the molecular clouds, and their parental cloud had possibly dispersed. 

\paragraph*{HD\,190833 Group.}
We identified 46 candidate members of the group. Similarly to the Cepheus Association, these sources are older and are not associated with cloud. Four of them show infrared excesses, characteristic of Class~II YSOs in the \wise\ wavelength region. The two subgroups, revealed by the \textit{W\/} distribution in Fig.\,\ref{fig:hdgroup_tanl}, are apparent also in Fig.\,\ref{fig:hdgroup}a and Fig.\,\ref{fig:hdgroup}d. Their mean distances and velocities are listed in Table~\ref{tab:subgroup_data}. Figure~\ref{fig:hdgroup}e shows that both subgroups are coeval, about 10--12~Myr old. The presence of some 10\,\% disc-bearing stars is noteworthy at this age.

\subsection{Young stars beyond 500 pc}

Our search for new candidate members of the Cepheus flare clouds was not extended beyond 500\,pc. Table~\ref{tab:yso_initial}, however, shows that a few young stars, projected onto the molecular clouds, are background stars around 900--1000\,pc. Massive \ion{H}{i} and CO clouds, associated with the Local Arm of the Galaxy are present in the high-latitude regions of Cepheus \citep[][respectively]{Heiles1967,Grenier1989} in this distance interval. The YSO candidate \spitzer\ sources around 1\,kpc,  projected onto the L1148/L1177 complex (see the black symbols in the lower panel of Fig.~\ref{fig1}) may be associated with this cloud complex. 

Another distant YSO, apparently not associated with known molecular cloud is the southern component of the visual double [K98c]\,EM*119. Its components, [KBK2009b]\,Em*\,119\,S and [KBK2009b]\,Em*\,119\,N are separated by 4\,arcsec, and both stars exhibit classical T~Tauri spectra \citep{kun2009}. The northern component has a distance of 327$^{+34}_{-26}$ \,pc, however the large RUWE value renders its \gaia~EDR3 data uncertain. The distance of the southern component is 907$^{+29}_{-23}$\,pc, with $\mathrm{RUWE} = 1.1$. The 3D extinction map of \citet{green2019} at the position of this star shows that the low extinction of $E(g-r)=0.14$\,mag  rises to  0.68\,mag around 900~pc.  
A similar visual double is [TNK2005]\,37 is found in the same part of the studied region. The brighter component [TNK2005]\,37\,c1 is a weak-line T~Tauri star \citep{TNK2005} at 974$^{+8}_{-9}$\,pc. 

\begin{table*}
\begin{center}	    	    
\caption{Average parallaxes, distances (from \citet{BJ2021}), and tangential velocities with the standard errors of the mean, and standard deviations ($\sigma$) for different substructures.}	\label{tab:result}	    	    
\begin{tabular}{lcccccccccccccc} 
\hline	    	    
Cloud &	N$^{*}$ & $\mathrm{N_{cand}}$ & $\mathrm{\varpi_{avg}}$ & $\mathrm{\sigma}_{\varpi}$ & $\mathrm{d_{avg}}$ & $\mathrm{\sigma}_\mathrm{d}$ & $\mathrm{v_{l,lsr,avg}}$ & $\mathrm{\sigma}_\mathrm{v_{l}}$ & $\mathrm{v_{b,lsr,avg}}$ & $\mathrm{\sigma}_\mathrm{v_{b}}$ & $\mathrm{L_{avg}\,(MST)}$\\
      & & & \multicolumn{2}{c}{(mas)} & \multicolumn{2}{c}{(pc)} & \multicolumn{4}{c}{(km s$^{-1}$)} & (pc)\\
\hline
NGC\,7023 & 50 & 31 & $2.906\pm0.016$ & 0.112 & $341\pm2$ & 13 & $-8.07\pm0.12$ & 0.84 & $-6.00\pm0.11$ & 0.81 & $0.540$ \\
L1148/L1158 & 3 & 0 & $2.978\pm0.011$ & 0.024 & $330\pm1$ & 3 & $-8.73\pm0.24$ & 0.51 & $-7.48\pm0.08$ & 0.17 & $3.261$ \\
L1177 & 42 & 40 & $2.911\pm0.010$ & 0.063 & $340\pm1$ & 8 & $-6.22\pm0.09$ & 0.60 & $-5.74\pm0.08$ & 0.51 & $1.087$ \\
L1217/L1219 & 39 & 37 & $2.719\pm0.016$ & 0.099 & $363\pm2$ & 13 & $-6.39\pm0.11$ & 0.70 & $1.63\pm0.09$ & 0.54 & $0.737$ \\
L1228 & 46 & 36 & $2.684\pm0.007$ & 0.050 & $368\pm1$ & 7 & $-4.27\pm0.16$ & 1.08 & $0.42\pm0.16$ & 1.08 & $1.176$ \\
L1235 & 71 & 68 & $2.930\pm0.008$ & 0.069 & $338\pm1$ & 8 & $-6.51\pm0.06$ & 0.50 & $2.45\pm0.05$ & 0.43 & $0.597$ \\
L1251 & 68 & 54 & $2.915\pm0.008$ & 0.063 & $339\pm1$ & 8 & $-4.66\pm0.10$ & 0.83 & $0.70\pm0.08$ & 0.65 & $0.619$ \\
Cepheus Ass. & 53 & 37 & $6.214\pm0.063$ & 0.463 & $161\pm2$ & 13 & $0.29\pm0.08$ & 0.58 & $2.34\pm0.08$ & 0.61 & $1.883$ \\
HD 190833 g. & 46 & 45 & $5.543\pm0.036$ & 0.250 & $180\pm1$ & 8 & $-2.24\pm0.08$ & 0.56 & $1.60\pm0.17$ & 1.15 & $1.690$ \\
\hline
\end{tabular}
\end{center}
\smallskip
\flushleft{\small
$^*$Total number of group members with reliable astrometric data in \gaia~EDR3.}
\end{table*}

\section{Overall view of the Cepheus flare region}
\label{sect5}

Table~\ref{tab:result} shows that the YSOs in the Cepheus flare are located between 330 and 370\,pc. The star-forming clouds in the $12\degr \loa b \loa 16\degr$ Galactic latitude interval, L1147/L1158, L1172/L1174, L1177, L1235, and L1251, are between 330 and 341\,pc. The clouds outside this latitude interval, L1217/L1219 and L1228 are 20--30\,pc farther. These results indicate a line-of-sight dimension of the cloud complex significantly smaller than its $\sim 90\times60$\,pc size, perpendicular to the line of sight. This is in accordance with the results of \citet{Heiles1967}, who found similar, sheet-like structure in the \ion{H}{i} counterpart of the cloud complex.

\begin{figure}
\centering \includegraphics[width=\columnwidth]{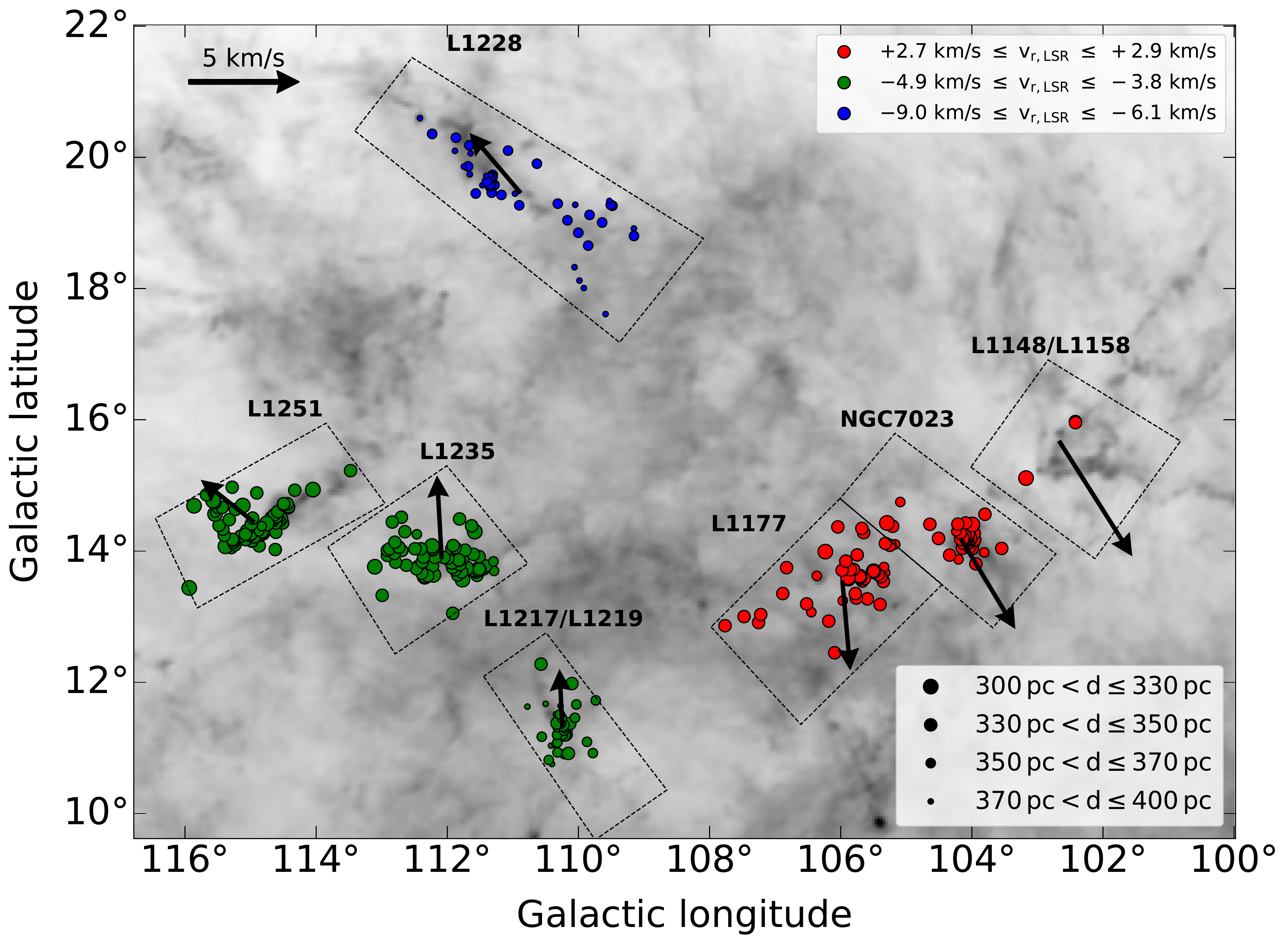}
\caption{Average tangential velocities of the stars associated with molecular clouds compared to the average of all known YSOs between 300 and 400\,pc. The colours indicate the radial velocities of the parental cloud \citep{YDM1997}, with respect to the LSR.}
\label{fig:vel_average}
\end{figure}

Figure~\ref{fig:vel} suggests that the large velocity dispersion reported by \citet{Dzib2018} results from the kinematic differences between the clouds. The substructures of the region also differ from each other in the radial velocity of the molecular gas \citep{YDM1997}. Radial velocities are available only for a few stars, and their uncertainties hinder any conclusion. To get an insight into the internal motions of the complex we assume that the average velocities of the YSOs and their natal clouds are similar and combine the tangential velocities of the stars with radial velocities of the clouds. The symbol colours in Fig.~\ref{fig:vel_average} indicate three distinct radial velocity intervals of the molecular clouds. Arrows indicate the mean tangential velocities of the young stellar groups associated with the clouds compared to the average of the whole system. Symbol sizes indicate distances.

The tangential and radial velocities reveal three kinematic groups. 
The clouds in the western part of the complex, L1148/L1158, L1172/L1174, and L1177 form a spatially and kinematically coherent structure.  Their star-forming histories are probably connected. An age gradient of the comoving stars can be seen in the L1148/L1177 sytem. Whereas no discless star and several embedded YSOs, not detected by \gaia, are found in the region of L1147/L1148,  90\,\% of the comoving stars around L1177 are discless low-mass stars.  This large structure overlaps in projection with the slightly more distant cluster \textit{Theia~5\/} identified by \citet{Kounkel2019}. 

The second, apparently kinematically coherent subsystem consists of L1219, L1235, and L1251. The tangential velocities of the comoving stars in L1219 and L1235 are similar, but L1219 is some 25\,pc farther than L1235. Star formation in L1219 was probably influenced by the supershell G113+10 from the southern direction \citep*{Bally2001,KMT2004}. The morphology of the region suggests that the star formation in L1235 might have been affected by the same supershell. 
L1251 is close to L1235 and both clouds have similar radial velocities, but their star-forming histories appear quite  different. The cometary shape of L1251 is indicative of star formation triggered by shock waves from the CFS \citep{TW1996,Olano2006}. The conspicuous cluster of stars, located on the head side of the cloud, and the apparent age gradient of YSOs observed across L1251 support this scenario. 

The third kinematic group of the region is L1228 at the highest Galactic latitude, at the distance of 368\,pc, and with the radial velocity of $-7.6$\,kms. Several arguments suggest that star formation in L1228 was also affected by the Cepheus Flare Shell \citep{Kirk2009}. L1228 is some 30\,pc farther than L1251, thus we can have a glimpse into the three-dimensional structure of the CFS. Whereas in the case of L1251 the shock from the supershell propagates in the east--west direction, largely perpendicular to the line of sight, L1228 and the adjacent [YDM97]\,CO\,60 are located towards the far side of the shell, where the expansion has a line-of-sight cmponent. The wide range of distances of the YSOs projected around L1228 in Table~\ref{tab:yso_initial} supports the scenario that star formation propagates largely along the line of sight, thus stars of different ages may be located at different distances.

The MSTs show that small, compact clusters are associated with NGC\,7023, L1219, L1228, L1235, and L1251, whereas loose aggregates of a few stars are associated with the L1148/L1158 complex and L1177. The comoving groups do not fill the defined tetragons, confirming that they are real clusters associated with the central molecular clouds.  We examined the cumulative distribution of the MST branch lengths, and determined the average length following the method described by \citet{Gutermuth2009}. The results are shown in the last column of Table~\ref{tab:result}. The average separations of the group members are 5--10 times larger than those of the young clusters observed by \spitzer, reflecting the more evolved nature of our groups.

The colour--magnitude diagrams suggest ages of 1--5 million years. Spectroscopic follow-up observations are essential for precise age determination.  Whereas 85~per cent of the stars in our initial list are disc-bearing YSOs, 87\,\% of the new candidate group members represent the discless population of the clouds. 

The Cepheus flare star-forming region is comparable in size and stellar content to the Taurus and Lupus clouds, however differs from them in structure and thus probably in star-forming history. The most striking difference is that whereas the Taurus clouds closely cluster around the central ridge of the cloud complex, and only a few smaller clouds are found in the outskirts of the region, star formation in the Cepheus flare occurs at the edges of the cloud complex. No young stars can be found in the central regions. Furthermore the depth of the Taurus complex is comparable with its projected size \citep{Galli2019}, whereas the Cepheus flare has a sheet-like shape. Unlike the YSOs of the Lupus clouds \citep{Galli2020}, stars in the Cepheus clouds form several kinematic subsystems. The overall picture of the region in view of the \gaia\ data confirms the scenario that the strucure of the cloud complex and star formation was probably shaped by shocks from energetic stellar winds and multiple supernovae. 

The distance of 160--180\,pc of the nearby groups of pre-main-sequence stars suggests their association with the wall of the Local Bubble \citep[e.g][]{Lallement2003}. The boundaries of the Local Bubble are defined by interstellar clouds, showing up as extinction of the starlight or interstellar absorption lines in stellar spectra. Figure~\ref{fig:Cepheus_Association}e suggests a small amount of reddening of the candidate pre-main-sequence association members. A colour excess $E(BP-RP)=0.205$\,mag, listed for HD\,190833 in the \gaia~DR2, also indicates a thin cloud toward the line of sight of this star.  The mean ages of the groups are compatible with the age of the Local Bubble \citep[14.6\,Myr,][]{Breit2006}. 
\section{Conclusions}
\label{sect6}
We studied the distribution and kinematics of the stars in the region $100\degr < l < 125\degr$, $8\degr < b < 22\degr$, based on the \gaia~EDR3 data. The area studied contains the molecular cloud complex of the  Cepheus flare and the adjacent cloud-free region. We identified \gaia~EDR3 counterparts of 176 stars, classified as YSOs in the literature. 

Based on the distances and tangential velocities of the known YSOs we found 266 additional candidate members of the molecular clouds. 
The colour--magnitude diagrams of the YSO groups suggest ages of 1--5 million years. 85\,\% of the stars in the initial list are disc-bearing YSOs, and 87\,\% of the new candidate group members represent the discless population of the clouds. 

The star-forming clouds in the $12\degr \loa b \loa 16\degr$ Galactic latitude interval, L1147/L1158, L1172/L1174, L1177, L1235, and L1251, have distances between 330 and 341\,pc. The clouds outside this latitude interval, L1217/L1219 and L1228 are 20--30\,pc farther. These distances suggest that the line-of-sight dimension of the cloud complex is significantly smaller than its $\sim 90\times60$\,pc size, perpendicular to the line of sight. 

We found that the clouds of the Cepheus flare form  three major kinematic subsystems: (i) L1147/L1148, L1172/L1174, and L1177, (ii) L1219, L1235, and L1251, (iii) L1228.  

We identified two kinematically distinct subgroups of candidate PMS stars around L1228. The southern subgroup is associated with a small and diffuse molecular cloud [YDM97]\,CO\,60. 

The 71 comoving stars associated with L1235, including the B8 type stars HD\,210806 and BD\,+72\degr1018 and the Herbig~Ae star SV\,Cep suggest the richest cluster of the Cepheus flare. 

We found a cluster of comoving stars, located on the head side of the cometary-shaped L1251. Comparison of the disc fractions of the off-cloud and on-cloud groups suggests an age gradient which fits into the scenario of triggered star formation. 

The \gaia\ parallaxes reveal a foreground population of pre-main-sequence stars, probably associated with the boundaries of the Local Bubble, and a few background stars, probably associated with the clouds of the Galactic Local Arm. 

We identified 37 new candidate members of the nearby Cepheus Association, at a distance of 161\,pc.

A new moving group of 46 candidate pre-main-sequence stars were identified around HD\,190833, at a mean distance of 180$\pm$1\,pc. 

\section*{Acknowledgements}
We thank to Adrienn Forr\'o for the helpful discussions.
This work has made use of data from the European Space Agency (ESA) mission
{\it Gaia} (\url{https://www.cosmos.esa.int/gaia}), processed by the {\it Gaia}
Data Processing and Analysis Consortium (DPAC,
\url{https://www.cosmos.esa.int/web/gaia/dpac/consortium}). Funding for the DPAC
has been provided by national institutions, in particular the institutions
participating in the {\it Gaia} Multilateral Agreement.
This work was supported by the ESA PRODEX Contract nr. 4000129910.
This research made use of Astropy,\footnote{http://www.astropy.org} a community-developed core Python package for Astronomy \citep{astropyI, astropyII}. For this work we have used Matplotlib \citep{matplotlib}, Pandas \citep{pandas}, astroquery \citep{astroquery}, TOPCAT \citep{topcat}, mst\_clustering \citep{MST_VanderPlas2016}, dustmaps \citep{dustmaps}.

\section*{Data availability}
The data underlying this article are available at Vizier at \url{https://vizier.u-strasbg.fr/viz-bin/VizieR}.
The datasets were derived from sources in the public domain of \textit{Gaia} at \url{https://gea.esac.esa.int/archive/}.

\bibliographystyle{mnras}
\bibliography{cepflare}
\bsp	% typesetting comment

\appendix
\section{Lists of young stellar objects of the Cepheus flare published in \gaia~EDR3}
%\clearpage
%\onecolumn
%\begin{landscape}
\renewcommand{\arraystretch}{1.5}

\begin{table}
\caption{Sample of YSOs contained in our initial list. The full table is available as supplementary material.}\label{tab:yso_initial}
\rotatebox{90}{
\begin{tabular}{lllrrrrrrr}
\hline
\multicolumn{1}{c}{Name} & \multicolumn{1}{c}{Gaia EDR3 ID} & \multicolumn{1}{c}{2MASS designation} & \multicolumn{1}{c}{RA (J2016)} & \multicolumn{1}{c}{DEC (J2016)} & \multicolumn{1}{r}{$\mu_{\alpha}^\star$} & \multicolumn{1}{c}{$\mu_{\delta}$} & \multicolumn{1}{c}{$\varpi$} & \multicolumn{1}{c}{Distance$^*$} & \multicolumn{1}{c}{RUWE}\\
& & & \multicolumn{2}{c}{(deg)}  & \multicolumn{2}{c}{(mas\,yr$^{-1}$)} & \multicolumn{1}{c}{(mas)} & \multicolumn{1}{c}{(pc)}\\
\hline 
2MASS J00003379+7940362 & 564638120384153472 & 00003379+7940362 & 0.1414 & 79.6767 & $22.753$ & $0.122$ & $6.456\pm0.025$ & $153.9_{-0.5}^{+0.5}$ & 0.993\\
BD+78 853 & 564638150446547200 & 00004121+7940398 & 0.1723 & 79.6778 & $23.013$ & $0.789$ & $6.463\pm0.014$ & $154.3_{-0.4}^{+0.3}$ & 0.917\\\relax
2MASS J00134052+7702104 & 540216042986714368 & 00134052+7702104 & 3.4182 & 77.0361 & $20.177$ & $-1.652$ & $5.887\pm0.029$ & $169.1_{-0.8}^{+0.8}$ & 1.697\\
TYC 4496-780-1 & 540216042984524800 & 00134052+7702104 & 3.4195 & 77.0364 & $21.892$ & $-1.608$ & $5.883\pm0.012$ & $169.5_{-0.3}^{+0.3}$ & 0.970\\
TYC4500-616-1 & 564977079203963648 & 00172828+7947580 & 4.3684 & 79.7994 & $22.244$ & $-1.105$ & $6.330\pm0.019$ & $157.5_{-0.4}^{+0.5}$ & 1.547\\\relax
2MASS J00380313+7903194 & 564698451789359104 & 00380313+7903194 & 9.5137 & 79.0554 & $24.969$ & $-4.077$ & $6.726\pm0.016$ & $148.3_{-0.4}^{+0.4}$ & 1.291\\\relax
[TNK2005] 4c1 & 564698451788613376 & 00380610+7903206 & 9.5260 & 79.0557 & $18.500$ & $-2.339$ & $6.817\pm0.048$ & $146.5_{-1.0}^{+1.3}$ & 4.070\\\relax
2MASS J00390355+7919191 & 564707973733132032 & 00390355+7919191 & 9.7654 & 79.3220 & $23.633$ & $-3.009$ & $6.583\pm0.023$ & $151.6_{-0.5}^{+0.6}$ & 1.739\\\relax
2MASS J00390430+7922350 & 564802050695978752 & 00390430+7922350 & 9.7685 & 79.3764 & $23.771$ & $-3.519$ & $6.751\pm0.066$ & $147.0_{-1.3}^{+1.4}$ & 1.088\\\relax
[TNK2005] 5c1 & 564707973733132800 & 00390619+7919096 & 9.7764 & 79.3193 & $24.209$ & $-2.477$ & $6.584\pm0.118$ & $152.2_{-2.5}^{+2.2}$ & 8.809\\\relax
2MASS J00393878+7905160 & 564695879105289344 & 00393878+7905160 & 9.9122 & 79.0878 & $24.510$ & $-2.834$ & $6.667\pm0.036$ & $149.0_{-0.9}^{+0.8}$ & 1.008\\
\hline
\end{tabular}}
%\smallskip
\flushleft{\small
$^*$Distance from \cite{BJ2021}}
\end{table}
%\end{landscape}

\begin{table}
\begin{center}
\caption{Column description of the table containing the YSOs and candidates. The list is available as supplementary material. The stars of the close groups and of the clouds are separated into two tables.}
\label{tab:yso-cand}
\begin{tabular}{ll}
\hline	    
Column & Description \\
\hline
cloud & Group identifier \\ 
Name & Star-identifier \\
designation & Gaia EDR3 identifier \\
TMASS & 2MASS identifier \\
ra & RA at J2016 \\
dec & DEC at J2016 \\
l & Galactic longitude \\
b & Galactic latitude \\
r\_med\_geo & Distance from \citet{BJ2021} \\
r\_lo\_geo & 16th percentile of distance posterior\\
& from \citet{BJ2021} \\
r\_hi\_geo & 84th percentile of distance posterior\\
& from \citet{BJ2021} \\
parallax & Parallax \\
parallax\_error & Error of the parallax \\
pmra & Proper motion in right ascension \\
pmra\_error & Error of the proper motion in right ascension \\
pmdec & Proper motion in declination \\
pmdec\_error & Error of the proper motion in declination \\
tang\_vel\_l & Tangential velocity in galactic longitude \\
tang\_vel\_b & Tangential velocity in galactic latitude \\
phot\_g\_mean\_mag & Magnitude in $G$ band \\
phot\_bp\_mean\_mag & Magnitude in $G_\mathrm{BP}$ band \\
phot\_rp\_mean\_mag & Magnitude in $G_\mathrm{RP}$ band \\
Jmag & Magnitude in 2MASS $J$ band \\
Hmag & Magnitude in 2MASS $H$ band \\
Kmag & Magnitude in 2MASS $Ks$ band \\
ruwe & Renormalized unit-weight error \\
\hline
\end{tabular}
\end{center}
\end{table}

\begin{table}
\begin{center}
\caption{Sample of the candidate disc-bearing YSOs classified by WISE/2MASS data. The full table is available as supplementary material.}
\label{tab:allwise}
\begin{tabular}{lll}
\hline
                         Name &           allwise\_id &               type \\
\hline
      2MASS J20144588+6942055 &  J201445.91+694205.5 &           Class II \\
      2MASS J20171018+7207013 &  J201710.24+720701.5 &           Class II \\
      2MASS J20525705+7245414 &  J205257.07+724541.5 &           Class II \\
      2MASS J20541245+7324249 &  J205412.49+732424.8 &           Class II \\
\hline
\end{tabular}

\end{center}
\end{table}

\end{document}